\documentclass{JHEP3}
\usepackage{amsmath,epsfig}

\newcommand{\be}{\begin{equation}}
\newcommand{\ee}{\end{equation}}
\newcommand{\bea}{\begin{eqnarray}}
\newcommand{\eea}{\end{eqnarray}}

\title{The minimal $3+2$ neutrino model versus oscillation anomalies}
\preprint{
	IFIC/12-34\\
	IFT-UAM/CSIC-12-47\\
	PPP/12/34
         DCPT/12/68\\
         EURONU-WP6-12-50\\
	}

\author{A. Donini$^{1,2}$, P.~Hern\'andez$^1$, J.~L\'opez-Pav\'on$^3$, M.~Maltoni$^2$ and T. Schwetz$^4$\\
       (1)  Instituto de F\'{\i}sica Corpuscular, CSIC-Universitat de Val\`encia\\
		Apartado de Correos 22085, E-46071 Valencia, Spain\\
		(2)Instituto de F\'{\i}sica Te\'orica UAM/CSIC, Calle Nicol\'as Cabrera 13-15, E-28049 Madrid, Spain \\
		(3)Institute for Particle Physics Phenomenology (IPPP), Department of Physics, Durham
University, Durham DH1 3LE, UK \\
(4) Max-Planck-Institut fuer Kernphysik, Saupfercheckweg 1, 69117 Heidelberg, Germany }

\abstract{
We study the constraints imposed by neutrino oscillation experiments on the minimal extension of the Standard Model that can explain neutrino masses, which requires the addition of just  two singlet Weyl fermions. The most general renormalizable couplings of this model imply generically four massive neutrino mass eigenstates while one remains massless: it is therefore a minimal $3+2$ model. The possibility to account for the confirmed solar, atmospheric and long-baseline oscillations, together with the LSND/MiniBooNE and reactor anomalies is addressed. We find that the minimal model can fit oscillation data including the anomalies better than the standard $3\nu$ model and 
similarly to  the $3+2$ phenomenological models, even though the number of free parameters is much smaller than in the latter. 
 Accounting for the anomalies in the minimal model  favours a normal hierarchy of the light states and requires a large  reactor angle, in agreement with recent measurements. Our analysis of the model employs a new parametrization of seesaw models that extends the Casas-Ibarra one to regimes 
 where higher order corrections in the light-heavy mixings  are significant. }

\begin{document}

\section{Introduction}

The simplest extension of the Standard Model that can account for the observed neutrino masses involves 
the addition of two extra singlet Weyl fermions. Such extension encompasses very different possibilities depending on the hierarchy of scales between the Dirac and Majorana masses. If the latter are zero the model corresponds to  two massive Dirac neutrinos; in the opposite limit when they are much larger than the electroweak scale, we get  the Type-I seesaw model  \cite{Minkowski:1977sc,GellMann:1980vs,Yanagida:1979as,Mohapatra:1979ia}. More exotic possibilities are the so-called  direct or inverse seesaws \cite{Wyler:1982dd,Mohapatra:1986bd} when the hierarchy is such that a global lepton number symmetry is  approximately preserved \cite{Gavela:2009cd}. 
Finally, if the Majorana mass scale is low but not zero one gets the so-called mini-seesaw models~\cite{deGouvea:2005er,deGouvea:2006gz}, that can be considered the minimal models with extra light sterile neutrinos \cite{Donini:2011jh}.

The most general renormalizable Lagrangian including two singlet Weyl fermions, compatible with the SM gauge symmetries is given by:
\begin{eqnarray}
{\cal L} = {\cal L}_{SM}- \sum_{\alpha,i} \bar l^\alpha_L Y^{\alpha i} \tilde\Phi \nu^i_R - \sum_{i,j} {1\over 2} \bar\nu^{ic}_R M_N^{ij} \nu_R^j+ h.c., \nonumber
\label{eq:lag}
\end{eqnarray}
where the Yukawa matrix $Y$ is a $3 \times 2$ matrix and $M_R$ is a symmetric matrix of dimension two. We can 
choose a basis where the neutrino mass matrix takes the form 
\begin{eqnarray}
{\mathcal M}_\nu = \left(\begin{array}{ll} 0 &m_Y\\
m_Y^T & M_N \end{array}\right),
\end{eqnarray}
where $M_N={\rm Diag}(M_1,M_2)$ is diagonal, while $m_Y$ is a $3\times 2$ complex matrix. 
It can be shown that generically the spectrum contains four massive states and one massless state. The mixing is described in terms of  four independent angles and three CP violating phases. 

It should be stressed that the number of free parameters in this model is much smaller than in the models usually referred
to as the $3+2$ model that has been recently revived as viable possibility to explain the LSND/MiniBooNE and reactor anomalies \cite{Peres:2000ic,Sorel:2003hf,Kopp:2011qd, Giunti:2011gz,Abazajian:2012ys}. 
 In those models a generic mass matrix of dimension $5$ is considered and therefore the ultraviolet completion involves  necessarily  
a more complex extension of the Standard Model. For example, one that involves heavier Majorana neutrinos some of which are integrated out, or several more Weyl 
species that pair up to make more than three  Dirac neutrinos. In the following we will refer to the $3+2$ models usually considered in the literature
as $3+2$ phenomenological model (PM), while the model with just two Weyl singlets is called $3+2$ minimal model (MM). Table~\ref{tab:param} summarizes the number of physical parameters that can affect oscillations of active neutrinos in the 3+2 PM and MM models compared with the standard $3\nu$ model.

\begin{table}
\begin{center}
\begin{tabular}{l|c|c|c}
\hline
Model &   $\#$ $\Delta m^2$ & $\#$ Angles & $\#$ Phases \\
\hline
3 $\nu$ &   2  & 3 & 1 \\
3+2 MM  &  4 &  4 & 3 \\
3+2 PM &   4 &  9 &  5 \\
\hline
\end{tabular}
\caption{Number of independent mass differences, mixing angles and CP phases (only those that can enter oscillations of active states) in the standard 3$\nu$ scenario, the 3+2 phenomenological model (PM) and the 3+2 minimal model (MM).  }
\label{tab:param}
\end{center}
\end{table}

In Ref.~\cite{Donini:2011jh},  the constraints of neutrino oscillation data on the $3+1$ MM and $3+2$ MM were studied. The former is completely excluded . For the latter, the parameter
space in the case of degenerate Majorana masses ($M_1=M_2=M$) was fully explored, and it was concluded that  only the regions 
$M \leq 10^{-9} (10^{-10})$ eV or $M \geq 0.6 (1.6)$ eV for the normal (inverted) hierarchy, respectively, are presently allowed by data. 
Furthermore, the degenerate limit with $M \sim$ eV did not improve the fits including  the LSND/MiniBooNE or new SBL reactor data in contrast with the $3+1$ PM model.
On the other hand, a qualitative analysis showed that the non-degenerate case with $M_1 \neq M_2 \sim O$(eV) could actually improve those fits, 
similarly to the $3+2$ PM.  The main objective of this paper is to quantify to what extent this is true. Analyses of the non-degenerate case have also been considered in \cite{deGouvea:2011zz, Fan:2012ca}. These studies have used the Casas-Ibarra parametrization that assumes an approximate decoupling of light and heavy sectors. For heavy masses in the $O$(eV) range, we find however that the Casas-Ibarra limit is not sufficiently precise and we have to use a more general parametrization.

We will not consider in this paper the limit $M_2 \gg M_1$ (3+1+1 minimal model), which corresponds to the proposal of \cite{Nelson:2010hz}, recently reconsidered in \cite{deGouvea:2011zz,Kuflik:2012sw}. 

The structure of the paper is as follows. In Sec.~\ref{sec:param} we introduce a new parametrization of seesaw-type models that extends
the one by Casas-Ibarra to a situation where the decoupling of light and heavy sectors is not  large enough.  In Sec.~\ref{sec:fits} we present the results of the global fit to the 3+2 MM and compare it with that
of the 3+2 PM \cite{Kopp:2011qd, Giunti:2011gz}, and with the standard 3$\nu$ model. In Sec.~\ref{sec:pheno} we show the present constraints in the 3+2 MM on the physical combinations that will be measured in  future neutrino oscillation experiments. In Sec.~\ref{sec:conclu} we conclude.

\section{Parametrization of seesaw models beyond Casas-Ibarra}
\label{sec:param}

There are many possible parametrizations of the mass matrix. A good choice will usually be one that satisfies two properties: 1) it contains
all independent parameters and no more, 2) it is convenient to impose existing constrains. Since two mass splittings have been determined
with high accuracy, ie. the solar and atmospheric one, it makes sense to use a parametrization that uses as parameters those physical quantities.

In the case when $m_Y \ll M_N$, a convenient parametrization that does precisely this is the one first introduced by Casas-Ibarra \cite{Casas:2001sr}, which exploits the approximate decoupling of the light and heavy sectors, using as parameters not only the  masses but also the mixings that have already been measured. 

If $M_i \sim O(eV)$, however, the Casas-Ibarra parametrization (which involves a perturbative expansion in $m_Y/M$) is not sufficiently precise. 
In fact, next order corrections in the $m_Y/M$ expansion are significant. In the following we present an alternative parametrization, inspired in Casas-Ibarra, 
 that coincides with it in the proper limit, but that does not assume any expansion in $m_Y$ or $M^{-1}_N$ and therefore can be used in the full parameter space. For an alternative parametrization with similar properties see \cite{Blennow:2011vn}.

There should be a $5\times 5$ unitary matrix such that 
\begin{eqnarray}
{\mathcal M}_\nu = U^*~ Diag(0,m_2,m_3,M_1,M_2)~ U^\dagger.
\end{eqnarray}
We can reduce one dimension by projecting out the zero mode, that is we easily find a $3\times 3$ unitary matrix $U_0$ such that
 ${\mathcal M}_\nu$ of the form
\begin{eqnarray}
\left(\begin{array}{ll} U_0^T & 0 \\
0 & I \end{array}\right) {\mathcal M}_\nu \left(\begin{array}{ll} U_0 & 0 \\
0 & I \end{array}\right) = \left(\begin{array}{ll} 0 & m_D \\
m_D^T & M \end{array}\right),
\end{eqnarray}
where now $m_D$ has zeros in the first row. A possible choice is to take the first row of $U_0^*$ 
of the form $\epsilon_{ijk} (m_Y)_{1i} (m_Y)_{2k}$ properly normalized. The remaining rows can be anything as long 
as it is unitary. 

In this way we reduce one dimension of the problem and find a $4\times 4$ unitary matrix, $V$, satisfying
\begin{eqnarray}
\left(\begin{array}{ll} 0 & m_D \\
m_D^T & M \end{array}\right)
 = \left(\begin{array}{ll} 1 & 0 \\
0 & V^* \end{array}\right) Diag(0,m_2,m_3,M_1,M_2) \left(\begin{array}{ll} 1 & 0 \\
0 & V \end{array}\right)^\dagger.
\label{eq:4d}
\end{eqnarray}


 Defining $m_l \equiv Diag(m_2,m_3)$ and $M_h\equiv Diag(M_1,M_2)$ and 
 \begin{eqnarray}
V\equiv \left(\begin{array}{ll} A & B \\
C & D \end{array}\right), 
\end{eqnarray}
in block form, eq.~(\ref{eq:4d}) implies
\begin{eqnarray}
A^* m_l A^\dagger + B^* M_h B^\dagger = 0 \rightarrow  I = - M_h^{-1/2} (B^*)^{-1} (A^*) m_l
A^\dagger (B^\dagger)^{-1} M_h^{-1/2}.
\label{eq:0}
\end{eqnarray}

 Therefore \`a la Casas-Ibarra we can define a 2$\times 2$ orthogonal matrix
\begin{eqnarray}
R^T \equiv - i m_l^{1/2} A^\dagger (B^\dagger)^{-1} M_h^{-1/2} \, ,
\end{eqnarray}
which allows to rewrite $B$ in terms of $A$:
\begin{eqnarray}
B^\dagger = -i M_h^{-1/2}  R m_l^{1/2} A^\dagger.
\end{eqnarray}

The unitarity of $V$ implies:
\begin{eqnarray}
A A^\dagger+ B B^\dagger = I \rightarrow  \left( I + m_l^{1/2} R^\dagger M_h^{-1} R m_l^{1/2}\right) = A^{-1} (A^\dagger)^{-1} \, .
\end{eqnarray}
The key point is now to use the polar decomposition of $A$
\begin{eqnarray}
A= W H,  \;\;\;H^\dagger =H, \;\;\; ~W^\dagger W = I,  
\end{eqnarray}
in the previous equation to get
\begin{eqnarray}
H^{-2} = I + m_l^{1/2} R^\dagger M_h^{-1} R m_l^{1/2}.
\label{eq:h}
\end{eqnarray}

Using unitarity,  the blocks $C$ and $D$ can also be rewritten as:
\begin{eqnarray}
 C^\dagger = - A^{-1} B D^\dagger =  -i m_l^{1/2} R^\dagger M_h^{-1/2} D^\dagger,
\end{eqnarray}
and using the polar decomposition of $D= \overline{W} \overline{H}$ we find
\begin{eqnarray}
 \overline{H}^{-2} = I + M_h^{-1/2} R m_l R^\dagger M_h^{-1/2}.
 \label{eq:hbar}
\end{eqnarray}
It can be shown that $\overline{W}$ is unphysical because it can be reabsorbed in a rotation of the sterile states, so we can fix it to the identity. 

The full unitary matrix can then be written as
\begin{eqnarray}
U =  \left(\begin{array}{lll}  U_{aa} & U_{as} \\
U_{sa} & U_{ss} 
\end{array}\right), 
\label{eq:u5}
\end{eqnarray}
with
\begin{eqnarray}
U_{aa} &= U_{PMNS} \left(\begin{array}{ll} 1 & 0\\
0& H \end{array}\right), ~ U_{as} &= i U_{PMNS} \left( \begin{array}{l} 0\\
  H m_l^{1/2} R^\dagger M_h^{-1/2} \end{array}\right),\nonumber\\
  U_{sa} &=  i \left( \begin{array}{ll} 0 & \overline{H} M_h^{-1/2} R m_l^{1/2}  
 \end{array}\right), ~
U_{ss} &= \overline{H}. 
\label{eq:par}
\end{eqnarray}
where $U_{PMNS}$ is a generic $3\times 3$ unitary matrix resulting from a combination of $U_0$ and $W$, while $H$ and $\bar{H}$ are 
defined in eqs.~(\ref{eq:h}) and (\ref{eq:hbar}).

More concretely, the physical parameters are chosen to be: the four mass eigenstates, three angles and two phases in $U_{PMNS}$:
\begin{eqnarray}
U_{PMNS}=\left( \begin{array}{ccc} 1 & 0 & 0\\
0 &  c_{23} &  s_{23}\\
0 & -s_{23} &  c_{23}  
\end{array} \right)  \left( \begin{array}{ccc}  c_{13} &  0 & s_{13} e^{-i \delta}\\
0 & 1 & 0 \\
-s_{13} e^{i \delta}& 0 & c_{13}  
\end{array} \right)  \left( \begin{array}{ccc} c_{12} &  s_{12}& 0\\
-s_{12} & c_{12} & 0 \\
0 & 0 & 1\end{array} \right)   \left( \begin{array}{ccc} 1 &  0& 0\\
0& 1 & 0 \\
0 & 0 & e^{i \alpha}\end{array} \right),  
\end{eqnarray}
 and one complex angle in $R$:
 \begin{eqnarray}
 R=\left( \begin{array}{cc}  \cos(\theta_{45} + i \gamma_{45}) &   \sin(\theta_{45} + i \gamma_{45})\\
 -\sin(\theta_{45} + i \gamma_{45}) &   \cos(\theta_{45} + i \gamma_{45}) 
\end{array} \right). 
 \end{eqnarray}
We note that, even though $U_{PMNS}$ is unitary and we have used the same notation as for the standard unitary mixing matrix in the $3\nu$ model, the  mixing in the $3\times 3$ light sector depends also on $H$, which contains the expected non-unitarity effects. 
To leading-order in the seesaw limit, $H \simeq \overline{H}\simeq I $ and in this case the mixing matrix simplifies to the Casas-Ibarra parametrization as expected. 
\begin{eqnarray}
U \rightarrow  \left(\begin{array}{ll} U_{PMNS}& i U_{PMNS} m_\nu^{1/2} R^\dagger M_h^{-1/2} \\
i M_h^{-1/2} R m_\nu^{1/2} & I \end{array}\right),  
\end{eqnarray}
where $U_{PMNS}$ is the standard $3\nu$ mixing matrix. 

For the NH case, this is the standard parametrization. However for the IH, since the order of the mass eigenstates is unchanged, the parametrization of $U_{PMNS}$ is not the standard one: it differs from it by  the permutation $123\rightarrow 231$, which in particular implies that the role of the solar angle is played by the angle $\theta_{13}$ and the role of the reactor angle is controlled by $\theta_{12}$ (with $\theta_{12}\approx \pi/2$). 

It is straightforward to extend this general parametrization to the $n_R=3$ case.  

The physical range of the parameters can be chosen to be $\theta_{ij} \in [0,\pi/2]$, $\delta, \alpha \in [0, 2 \pi]$ and $\gamma_{45} \in (-\infty, \infty)$. In contrast with Casas-Ibarra, no divergence occurs for large $\gamma_{45}$ since the mixing matrix converges exponentially to a finite result when this parameter grows. 

\vspace{0.5cm}
The model can obviously be matched to a $3+2$ phenomenological model, with $\Delta m^2_{41}= M_1^2, \Delta m^2_{51}= M_2^2$,  that are free parameters, and with the $5\times 5$ unitary matrix given by eqs.~(\ref{eq:u5}) and ~(\ref{eq:par}). The elements $U_{e4}, U_{e5}, U_{\mu4}$ and $U_{\mu5}$ are well known functions of the chosen parameters, and are strongly correlated among themselves and with the masses.
In Ref.~\cite{Donini:2011jh, deGouvea:2011zz,Fan:2012ca}, the expression for these quantities in various limits was given using the Casas-Ibarra parametrization. 

The expressions for the more general parametrization beyond Casas-Ibarra limit are not as simple. We can derive, however, some robust results in various limits.
For the NH in the limit of vanishing solar mass splitting (i.e. $m_2=0$), the heavy-light mixings simplify significantly to 
\begin{eqnarray}
U_{e4} &=& i \sqrt{m_3 M_2 \over X} e^{i (\alpha-\delta)} s_{13} \sin z_{45}  + {\mathcal O}(\sqrt{m_2/M_i}) \, , \nonumber\\
U_{e5} &= & i \sqrt{m_3 M_1 \over X} e^{i (\alpha-\delta)} s_{13} \cos z_{45}+ {\mathcal O}(\sqrt{m_2/M_i}) \, , \nonumber\\
U_{\mu 4} &=  & i \sqrt{m_3 M_2 \over X} e^{i \alpha} c_{13} s_{23} \sin z_{45}+ {\mathcal O}(\sqrt{m_2/M_i})  \, , \nonumber\\
U_{\mu 5} &= & i \sqrt{m_3 MLBL 1 \over X} e^{i \alpha} c_{13} s_{23} \cos z_{45}+ {\mathcal O}(\sqrt{m_2/M_i}) \, .
\label{eq:usnh-a}
\end{eqnarray}
where $z_{45} \equiv \theta_{45} - i \gamma_{45}$ and
\be
X\equiv M_1 M_2 + m_3 (M_1 - M_2)/2 \cos (2 \theta_{45})  + m_3 (M_1 +M_2)/2 \cosh (2 \gamma_{45}) \, .
\ee
 It is easy to check that the quantity $\phi_{45} \equiv \arg(U_{e4}^* U_{e5} U_{\mu4}U_{\mu5}^*)$, which drives the CP violation  in the $\mu-e$ probability in the LSND/MiniBooNE range, is exactly zero in this limit.  

We also see that, generically, the electron couplings are suppressed by $\theta_{13}$.  
These couplings are those that get modified more significantly by the effects of the solar splitting. In the Casas-Ibarra limit we find the first correction to be 
\begin{eqnarray}
U_{e4} &\simeq& i \left(\sqrt{m_3\over M_1} e^{i (\alpha-\delta)} s_{13} \sin z_{45} + \sqrt{m_2\over M_1}  c_{13} s_{12} \cos z_{45} \right),\nonumber\\
U_{e5} &\simeq & i \left(\sqrt{m_3\over M_2} e^{i (\alpha-\delta)} s_{13} \cos z_{45}-  \sqrt{m_2\over M_2}  c_{13} s_{12} \sin z_{45} \right).
\label{eq:usnh-b}
\end{eqnarray}
CP violation in $\phi_{45}$ starts therefore at ${\mathcal O}(\sqrt{m_2} s_{13})$ or ${\mathcal O}(m_2)$. Both terms can be comparable in size. 
Based on these analytical results, it was argued that if the hierarchy is normal (NH),  the heavy-light mixings would be too small to reach the best fit points found in \cite{Kopp:2011qd, Giunti:2011gz}. We will see 
that this is not the case. The NH works thanks in part to the recent indication of a large {\it reactor} angle from Double CHOOZ, Daya Bay and RENO, and in part to the fact that actually corrections to the Casas-Ibarra limit are significant. 

The $\mu$ couplings, on the other hand, are not small for the NH. In order to keep them small enough, there must be also a partial cancellation with
 higher order corrections in eq.~(\ref{eq:usnh-a}). It is generic that if such a cancellation works for $\mu$, it becomes an enhancement in the
$\tau$ channel and viceversa.  

For the IH case, the limit of vanishing solar splitting is more complicated and the phase $\phi_{45}$ does not vanish in general.
In fact, even the limit of vanishing solar splitting and reactor angle (i.e. $m_2=m_3\simeq \sqrt{\Delta m^2_{atm}}, \theta_{12} = \pi/2$)  is non-trivial in this case.
In the Casas-Ibarra limit we get
\begin{eqnarray}
U_{e4} &\simeq& i \sqrt{m_3\over M_1} \left(e^{i (\alpha-\delta)} s_{13} \sin z_{45} +c_{13}  \cos z_{45} \right) \, ,\nonumber\\
U_{e5} &\simeq & i \sqrt{m_3\over M_2} \left(e^{i (\alpha-\delta)} s_{13} \cos z_{45}-    c_{13} \sin z_{45} \right) \, , \nonumber\\
U_{\mu4} &\simeq& i \sqrt{m_3\over M_1} e^{i \delta} s_{23} \left(e^{i (\alpha-\delta)}  c_{13}  \sin z_{45} - s_{13}  \cos z_{45} \right)\, , \nonumber\\
U_{\mu5} &\simeq & i \sqrt{m_3\over M_2} e^{i\delta} s_{23} \left(e^{i (\alpha-\delta)} c_{13} \cos z_{45}+   s_{13} \sin z_{45} \right)\, ,
\label{eq:usih}
\end{eqnarray}
where in the IH  case $s_{13}$ corresponds to the solar angle and is therefore not small. The mixings are in the right ballpark in this case
for $M_i \sim $ eV \cite{Donini:2011jh}. 


\section{Oscillation data and the $3+2$ minimal model}
\label{sec:fits}

As we have seen, the $3+2$ minimal model  can be represented as a $3+2$ PM, as the previous parametrization shows explicitely. On the other hand, the parameters in the mixing matrix are strongly correlated. The active-sterile mixing, for example, generically depends on all the angles and masses, light and heavy. 

The analysis of oscillation data in the context of the 3+2 PM \cite{Kopp:2011qd} is based on the sensible assumption that most of the parameters can be assigned to one of  two groups: the ones that  contribute mostly to  long-baseline (LBL)  ocillation data (including KamLAND, MINOS, CHOOZ, T2K, solar and atmospheric data), such as the two lighter mass splittings, and the ones that contribute mostly to short-baseline (SBL) data (short baseline reactors, LSND/MiniBooNE), such as the heavier mass splittings. Only the combinations $d_e\equiv |U_{e4}|^2+|U_{e5}|^2$ and $d_\mu=|U_{\mu4}|^2+|U_{\mu5}|^2$ can be significantly constrained by both \cite{Maltoni:2007zf}.  The LBL constraints on these combinations are added to the $\chi^2_{SBL}$ as pulls and from  the combined $\chi^2$, the heavy masses and the active-sterile mixings are extracted. For more details see Ref.~\cite{Maltoni:2007zf, Kopp:2011qd}. 

The decoupling of the LBL and SBL analyses does not hold so cleanly in the $3+2$ MM, because the number of independent parameters in the mixing matrix 
is significantly smaller and correlations are large.  
A global fit would require to deal with all parameters: however, this is too time consuming and therefore some approximations are needed. 
The parametrization proposed above reflects the approximate decoupling of the heavy sector when $M_i \gg m_i$ that can obviously be exploited. 
In particular, for NH, the parameters: $\theta_{12}, \theta_{23},  m_2, m_3$, which  dominantly drive the solar and atmospheric oscillation, are very well constrained by the LBL data\footnote{For IH the angle relevant for solar oscillations in $\theta_{13}$ instead of $\theta_{12}$.}. In order to fix them we will therefore minimize only $\chi^2_{LBL}$. The data samples we include as LBL are: KamLAND \cite{PhysRevD.83.052002}, MINOS-CC \cite{PhysRevLett.106.181801}, T2K \cite{Abe:2011sj}. In our SBL data sample we follow the same procedure as in \cite{Kopp:2011qd} and  include:   LSND \cite{Aguilar:2001ty}, MiniBooNE \cite{mb1,mb2}, KARMEN \cite{Armbruster:2002mp}, NOMAD  \cite{Astier:2001yj}, CDHS \cite{Dydak:1983zq}, as well as the reactor data Bugey3 \cite{bugey3}, Bugey4 \cite{bugey4}, ROVNO \cite{rovno}, Krasnoyarsk  \cite{kras}, G\"osgen  \cite{gos}, CHOOZ  \cite{Apollonio:2002gd} and Palo Verde \cite{paloverde}. Oscillations at these shorter baselines do depend also on the previous parameters  (see eqs.~(\ref{eq:usnh-a}-\ref{eq:usih})), but SBL data is less constraining, because of  larger uncertainties both theoretically (dependence on more free parameters) and experimentally.

On the contrary, the heavy masses $M_1$ and
$M_2$ are only relevant at short baselines. Therefore they can be determined by minimizing $\chi^2_{SBL}$. This has been already done in the context of the $3+2$ PM fits \cite{Kopp:2011qd, Giunti:2011gz}, and relatively small regions were found for these masses. We will 
therefore use the best fit points of those fits to fix $M_1, M_2$. The results of the best fit points we consider  are summarized in Table~\ref{tab:3+2}.  

\begin{table}
\begin{center}
\begin{tabular}{l|l|l|l|l|l|l|l}
\hline
$3+2$ PM & $|\Delta m^2_{41}|$(eV$^2$) &  $|\Delta m^2_{51}|$(eV$^2$) & $|U_{e4}|$ & $|U_{e5}|$ & $|U_{\mu4}|$ & $|U_{\mu5}|$& $\phi_{45}$\\
\hline
KMS\cite{Kopp:2011qd} & 0.47 & 0.87 & 0.128 & 0.138 & 0.165 & 0.148 & 1.64 $\pi$ \\
GL\cite{Giunti:2011gz} & 0.9 & 1.61 & 0.13 & 0.13  & 0.14  & 0.078 &1.51 $\pi$ \\
\hline
\end{tabular}
\end{center}
\caption{Best fit values of the heavy mass splittings in the $3+2$ phenomenological model fits from \cite{Kopp:2011qd, Giunti:2011gz}.} 
\label{tab:3+2}
\end{table}

The remaining five parameters $(\theta_{13}(\theta_{12}), \delta, \alpha, \theta_{45}, \gamma_{45})$  for NH(IH)  can affect significantly both LBL, SBL oscillations, and obviously also the reactor oscillation recently  discovered by Double CHOOZ \cite{Abe:2011fz}, Daya Bay \cite{An:2012eh} and RENO\cite{Ahn:2012nd}. We will use a combined $\chi^2 = \chi^2_{LBL}+ \chi^2_{SBL}+\chi^2_{\rm Daya Bay}+\chi^2_{\rm RENO}$ to fit them. 

Summarizing, the global $\chi^2$ is computed in a five-dimensional  grid, after minimizing $\chi^2_{LBL}$ over the parameters $m_2, m_3, \theta_{13}(\theta_{12}), \theta_{23}$ for  NH(IH) and fixing $M_1, M_2$ to their best fit values in Table~\ref{tab:3+2}.  The remaining free parameters are allowed to vary within their full physical ranges.

\begin{figure}[htbp]
\begin{center}
\includegraphics[width=7cm]{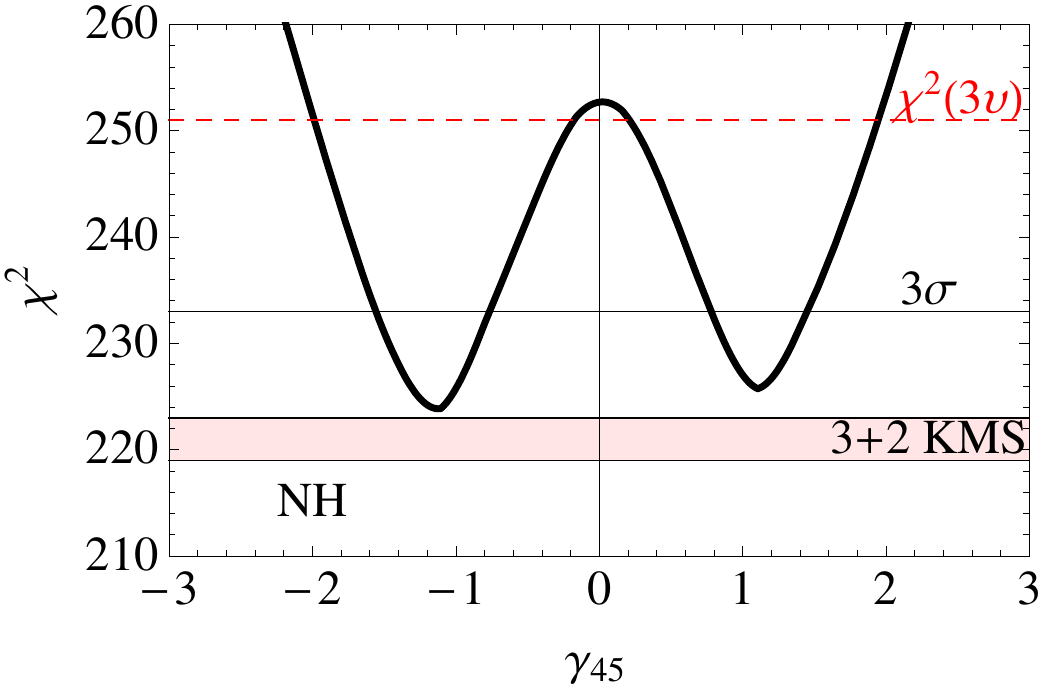}\includegraphics[width=7cm]{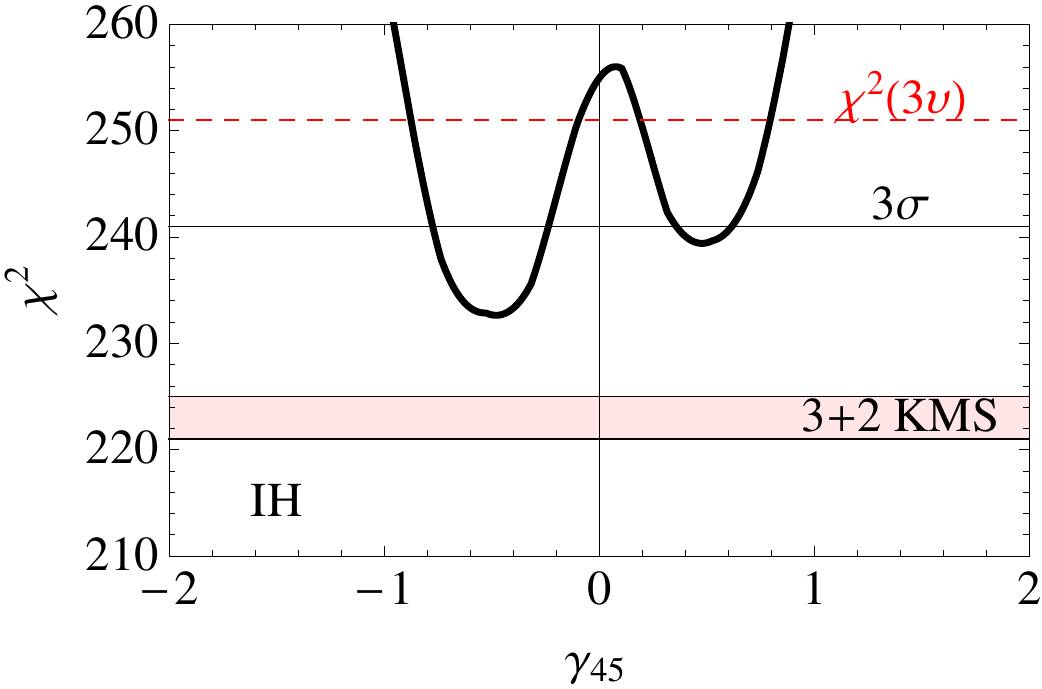}\\
\includegraphics[width=7cm]{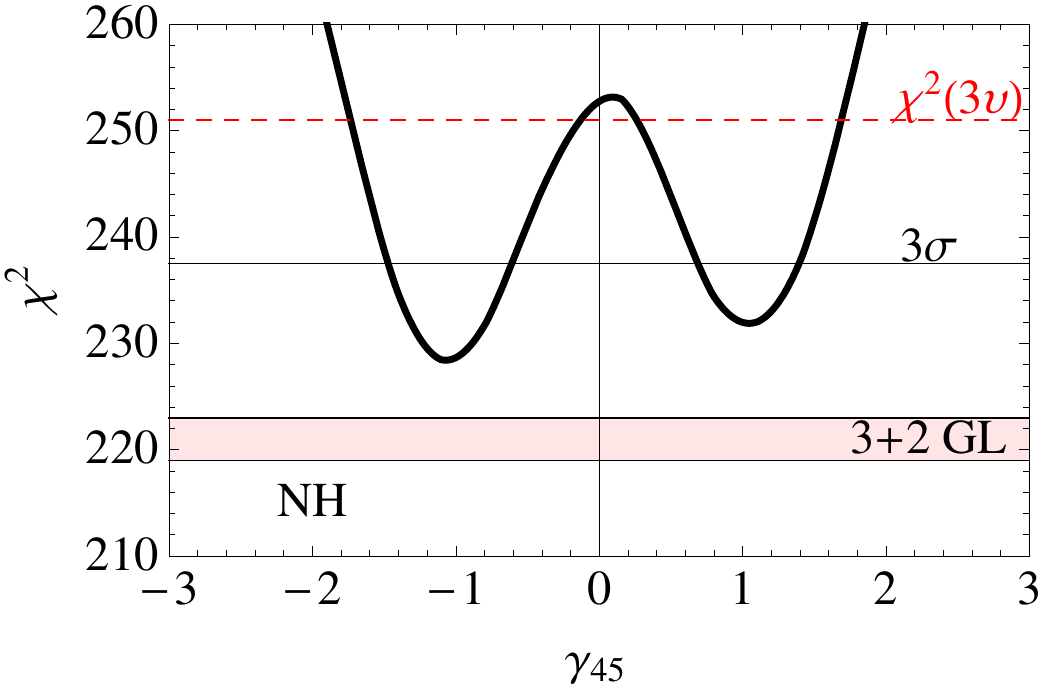}\includegraphics[width=7cm]{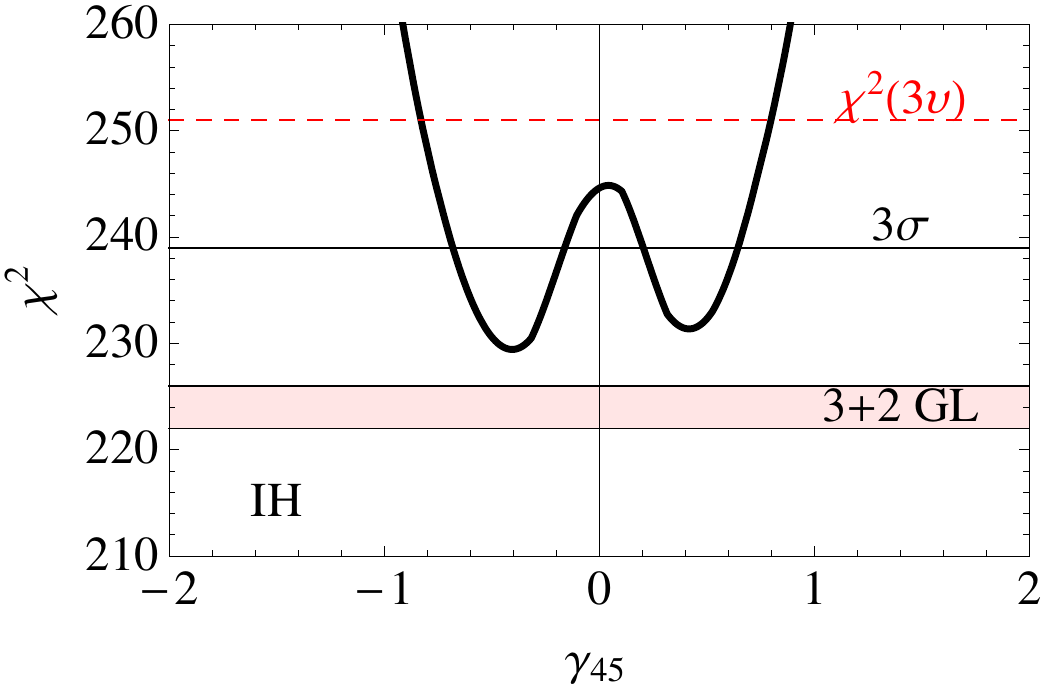}\\
\caption{ Minimum $\chi^2$ in the 3+2 MM  as a function of the parameter $ \gamma_{45}$ for NH (left) and IH (right), and for $M_1=\sqrt{\Delta m^2_{41}}$, $M_2=\sqrt{\Delta m^2_{51}}$ fixed to the best fit values of the 3+2 PM KMS fit (up) and to the GL fit (down). The dashed line corresponds to the 
minimum $\chi^2$ of the standard 3$\nu$ model. The band is the 2$\sigma$ band of the  3+2 PM KMS/GL best fit.}
\label{fig:chi2_g45}
\end{center}
\end{figure}

In Fig.~\ref{fig:chi2_g45} we show the  results, for the NH and IH,
of the one-dimensional projection of the total $\chi^2$  as a function of the complex angle $\gamma_{45}$, that is the parameter better constrained in both cases. The projections on the other parameters are  not particularly illuminating, because there are strong correlations. In the next section we will consider instead the constraints on several directly measurable combinations of the mixing matrix elements.  

As shown in Fig.~\ref{fig:chi2_g45}, the 3+2 MM outperforms the standard 3$\nu$ scenario, for both NH and IH, and for the two choices of $M_1$ and $M_2$ in Table~\ref{tab:3+2}. We get a  better fit for the KMS choice of $M_1$ and $M_2$ for NH, while the GL values of $M_1$ and $M_2$ give a slightly better fit in the IH case. The shift in the minimum $\chi^2$ for the two best  cases is  found to be
\begin{eqnarray}
\Delta \chi^2 = \chi^2_{min}({{\rm 3+2 ~MM}})  - \chi^2_{min}(3\nu)  \sim 28 ~(21)  \;\;\;{\rm NH(IH)}
\end{eqnarray}
which is significant given that the difference in the number of free parameters in both models is five (two of which are phases).  However, the improvement 
comes solely from the SBL data sample, while the fit to the LBL data deteriorates. This is reflecting the well-known tension between appearance and disappearance data. We find that  MINOS CC sample imposes very significant constraints. 
The reason is that the MINOS near detector is too near and  no significant effect of the heavier splittings is visible there, but the far detector CC sample is depleted at high energies due to the presence of the heavier states. The data do not show any indication of such depletion.

As an indication, we find that $\chi^2_{min} ({\rm LBL+SBL})- \chi^2_{min}({\rm LBL})  - \chi^2_{min}({\rm SBL})  \sim 16 (19)$ for NH (IH) for the KMS choice 
of $M_1, M_2$.


Compared to the  3+2 PM, the fit of the minimal model is necessarily worse, since it is equivalent to a $3+2$ PM with constraints.
 On the other hand Fig.~\ref{fig:chi2_g45} shows that the  fit does not degrade very significantly,  especially in the case of NH and the KMS choice of $M_1$ and $M_2$. This is non trivial because the 3+2 MM is  significantly more predictive. The difference in the number of physical parameters of  the 3+2 PM and 3+2 MM is 7 (5 angles and 2 phases). 
  
\begin{figure}[htbp]
\begin{center}
\includegraphics[width=7cm]{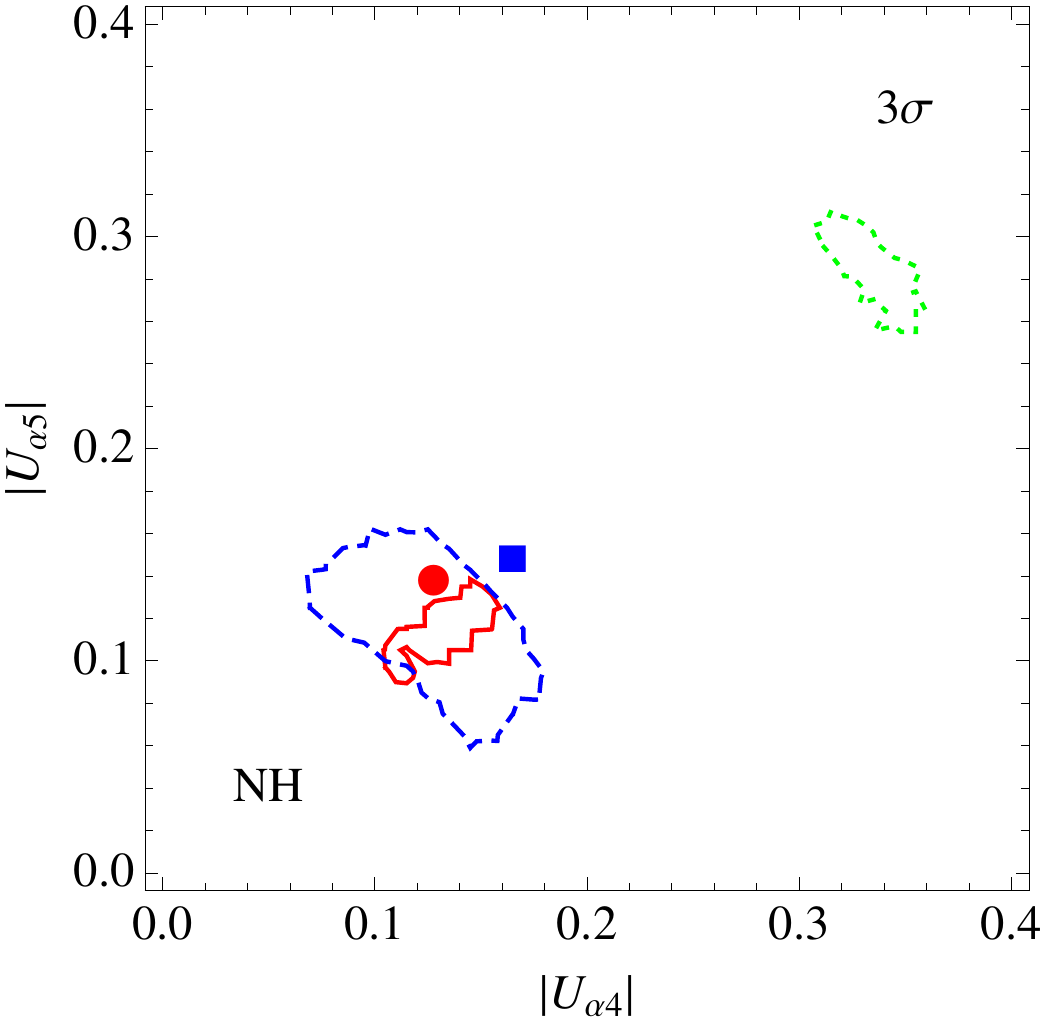}\includegraphics[width=7cm]{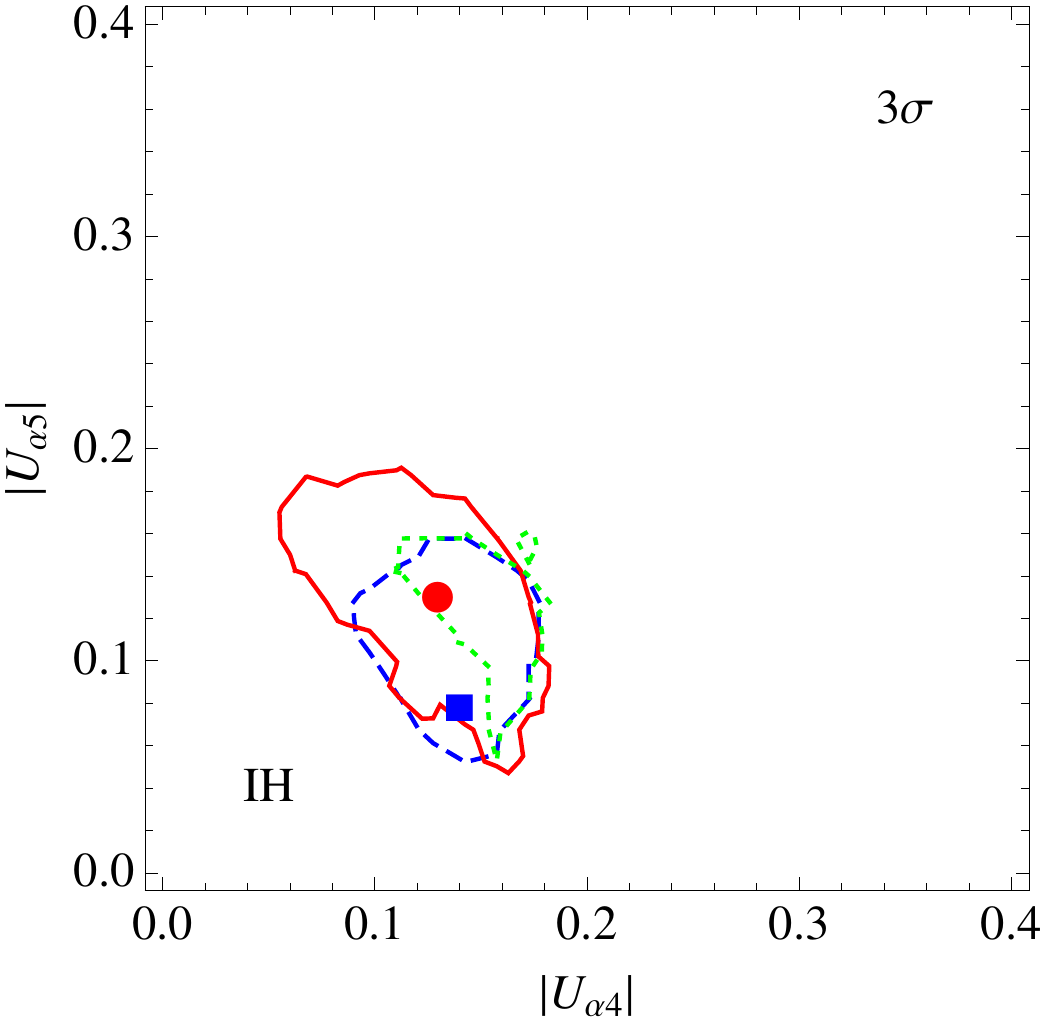}
\caption{Left: 3$\sigma$ ranges on the plane ($|U_{\alpha4}|$,$|U_{\alpha5}|$) for $\alpha=e$ (solid),$\mu$ (dashed) and $\tau$ (dotted) for the NH, with $M_1$ and $M_2$ fixed to the KMS values of Table~2.
Right: same for the IH and with the GL values of $M_1$ and $M_2$. The symbols: circle (e) and square ($\mu$)  
correspond to the best fit points of the 3+2 PM fits from Table~2.}
\label{fig:uax}
\end{center}
\end{figure}

\begin{figure}[htbp]
\begin{center}
\includegraphics[width=7cm]{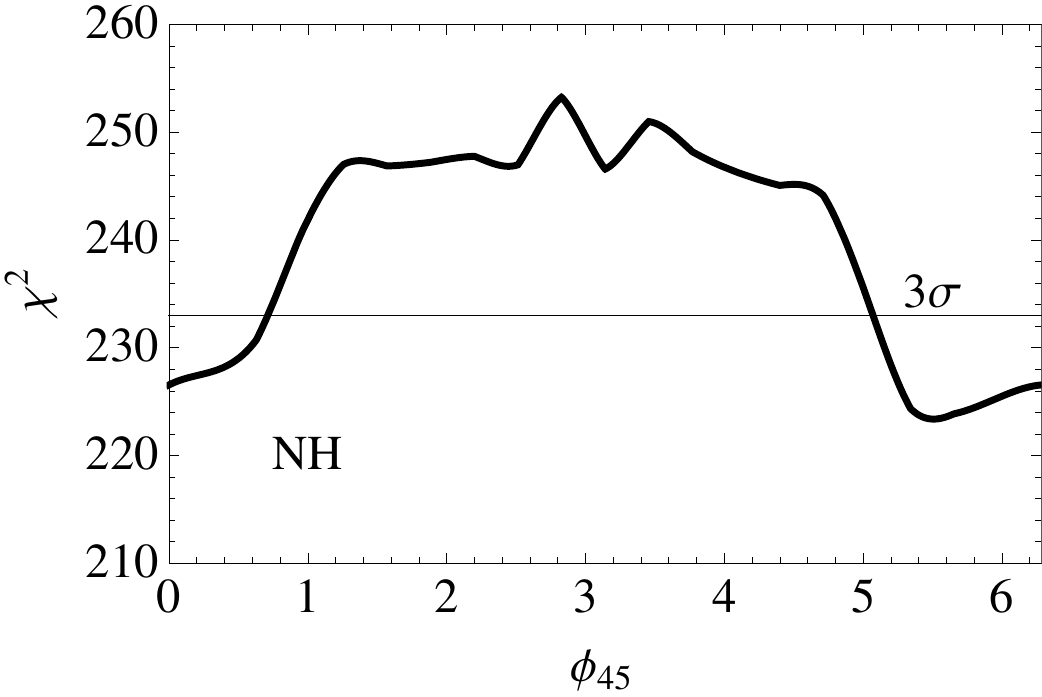}\includegraphics[width=7cm]{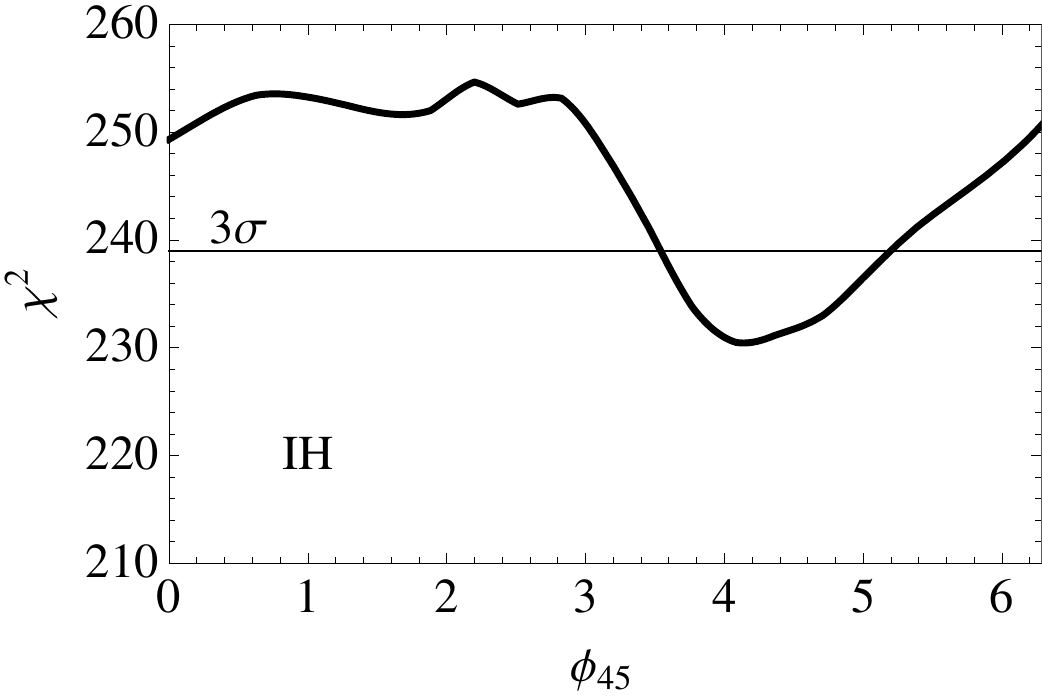}
\caption{Left: Projection of  the mimimum $\chi^2$ as a function of the SBL phase, $\phi_{45}$, for NH and with $M_1$ and $M_2$ fixed to the KMS values of Table~2. Right: same for IH and for $M_1$ and $M_2$ fixed to the GL values of Table~2.}
\label{fig:phi45}
\end{center}
\end{figure}

 In Figs.~\ref{fig:uax} we show the 3$\sigma$ limits on  the heavy-light mixing elements $|U_{\alpha i}|$, $i=4,5,\alpha=e,\mu,\tau$ in the 3+2 MM. The active-sterile couplings for the electron and muon sectors are in the same ballpark as those found in the phenomenological fits. The $\tau$ mixings, which are undetermined in the phenomenological fits, are very constrained in this case and turn out to be rather large for the NH.  In Fig.~\ref{fig:phi45} we show the limits on the SBL phase $\phi_{45} \equiv \arg(U_{e4}^* U_{e5} U_{\mu4}U_{\mu5}^*)$.  The values of the electron and muon mixing with the heavy states obtained for the 3+2 MM at the best fit point for NH (KMS masses) and IH (GL masses) are shown in Table~\ref{tab:3+2mm}.
 We note that the corrections to the Casas-Ibarra limit are rather significant, up to $30\%$ in some cases. 
\begin{table}
\begin{center}
\begin{tabular}{l|l|l|l|l|l|l|l}
\hline
$3+2$ MM & $|\Delta m^2_{41}|$(eV$^2$) &  $|\Delta m^2_{51}|$(eV$^2$) & $|U_{e4}|$ & $|U_{e5}|$ & $|U_{\mu4}|$ & $|U_{\mu5}|$& $\phi_{45}$\\
\hline
NH & 0.47 & 0.87 &  0.149 &  0.127 & 0.112 & 0.127 &1.8 $\pi$ \\
IH  & 0.9 & 1.61 & 0.139 & 0.122  & 0.138  & 0.107 &1.4 $\pi$ \\
\hline
\end{tabular}
\end{center}
\caption{Best fit values of the heavy mass splittings in the $3+2$ MM for NH and IH. The masses are not fitted, are fixed to the best fit points
of the phenomenological fits.}
\label{tab:3+2mm}
\end{table}

 One might worry that significant constraints can be obtained from
 the neutral current measurement in MINOS, which we have not included. We have checked that the value of the parameter $f_s$ of \cite{minosnc}, which measures the 
 sterile fraction of the muon disappearance,  is around 0.2 at the best fit, so it is well within the $90\%$ CL of the MINOS bound. 
    Future searches for  $\tau$ appearance oscillations however  could be useful to further constrain this mode \cite{deGouvea:2011zz}. One would need an experiment able to search for appearance of $\tau$'s, ie with sufficient high energy, and this would require an intermediate baseline
    of $L\sim O(10) $km. At the oscillation peak the $P_{\mu\tau}$ probability would be above $15\%$. At OPERA one would expect a constant contribution in the oscillation probability of $\sim 2.5\%$, which is probably too small to be detectable.

\section{Phenomenological implications for  LBL reactors and accelerators}
\label{sec:pheno}

The LBL reactor and T2K/NOVA measurements aim at the subleading atmospheric oscillation in the electron channel. In this section we review what to expect in the 3+2 MM model, since the previous fits constrain most of these observables. In a more general PM model, there is more freedom and some of these predictions can be relaxed. 

In reactor and accelerator experiments the far detector is located at a baseline comparable to solar or atmospheric oscillations and therefore, being the heavier
mass splittings significantly larger than the atmospheric one, are in the fast-oscillation regime. In appendix A, we give the general oscillation probabilities in this regime. Often these experiments also have near detectors to monitor the neutrino flux and further 
control other systematic errors. While the near dectector in reactor experiments are typically also rather far, so that the heavier splittings are still in the  fast oscillation regime, this is not the case with accelerator experiments, where the near detectors are at a distance where either 
oscillations have not taken place or they are about to. 

The electron disappearance probability measured by reactors, in the approximation of neglecting the solar 
mass splitting,  can be written as
\begin{eqnarray}
\left.P_{ee}\right|_{\rm reactor} \simeq N_{ee} \left[ 1 - A_{ee} \sin^2 \left({\Delta m^2_{ atm} L \over 4 E} \right)\right],
\end{eqnarray}
with 
\begin{eqnarray}
N_{ee} &\equiv& 1- 2 \left( |U_{e4}|^2 + |U_{e5}|^2 - |U_{e4}|^4-|U_{e5}|^4-|U_{e4}|^2 |U_{e5}|^2\right), \nonumber\\
A_{ee}  N_{ee} &\equiv& \left\{\begin{array}{lrrr} 
 4 |U_{e3}|^2 (|U_{e1}|^2 + |U_{e2}|^2),  & & & {\rm NH}\\
4 |U_{e1}|^2 (|U_{e3}|^2 + |U_{e2}|^2).  & & & {\rm IH} \end{array}\right. \\
\end{eqnarray}
Reactor experiments such as Daya Bay and RENO use near detectors to fix the normalization, therefore they are sensitive to the 
quantity $A_{ee}$.  Note that  the near detectors in both experiments are located sufficiently far to observe averaged oscillations for the two large mass splittings . 

\begin{figure}[htbp]
\begin{center}
\includegraphics[width=7cm]{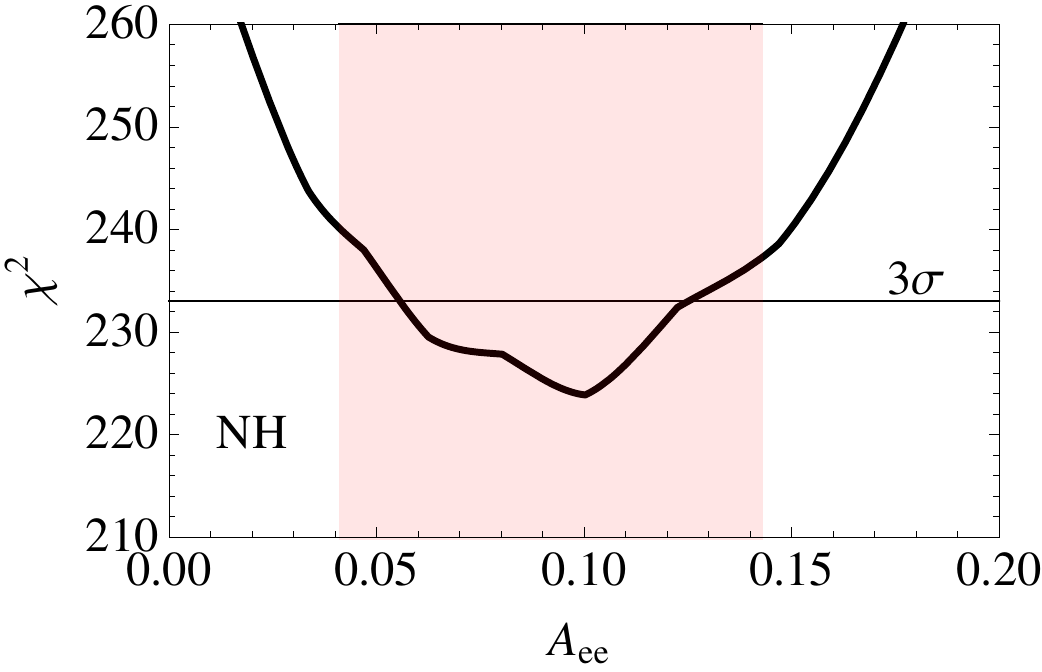}\includegraphics[width=6.7cm]{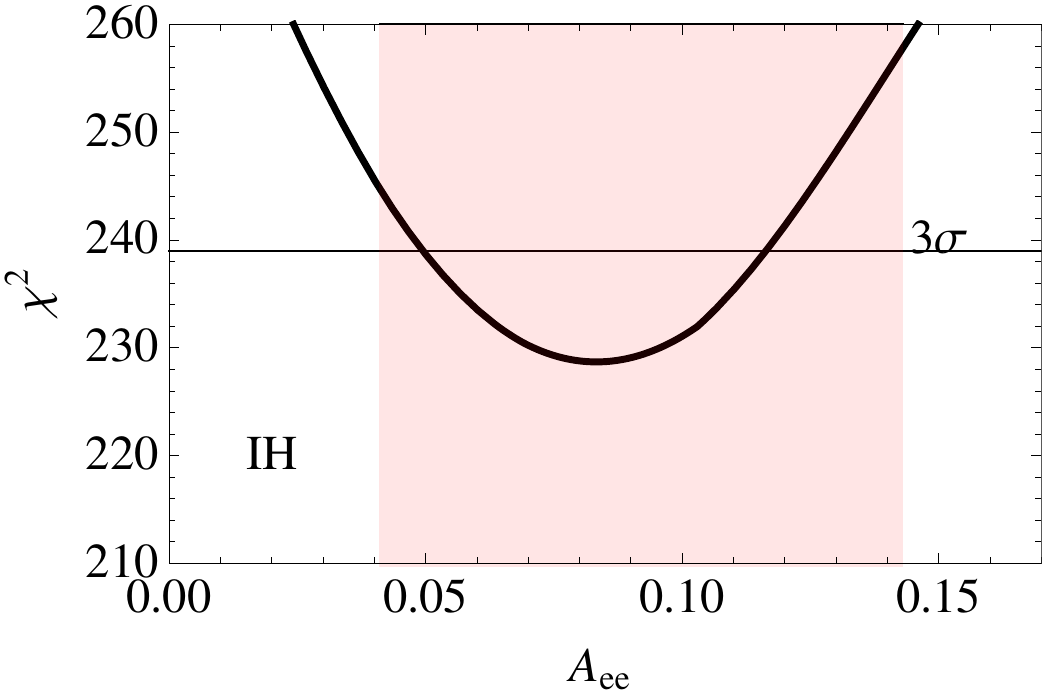}
\includegraphics[width=7cm]{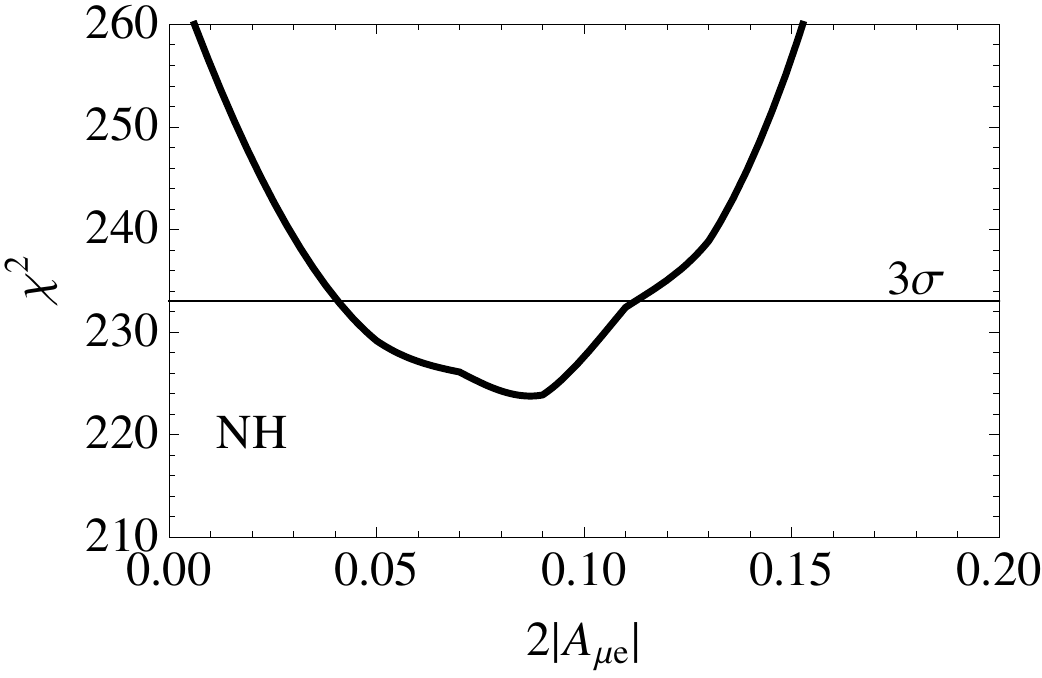}\includegraphics[width=7cm]{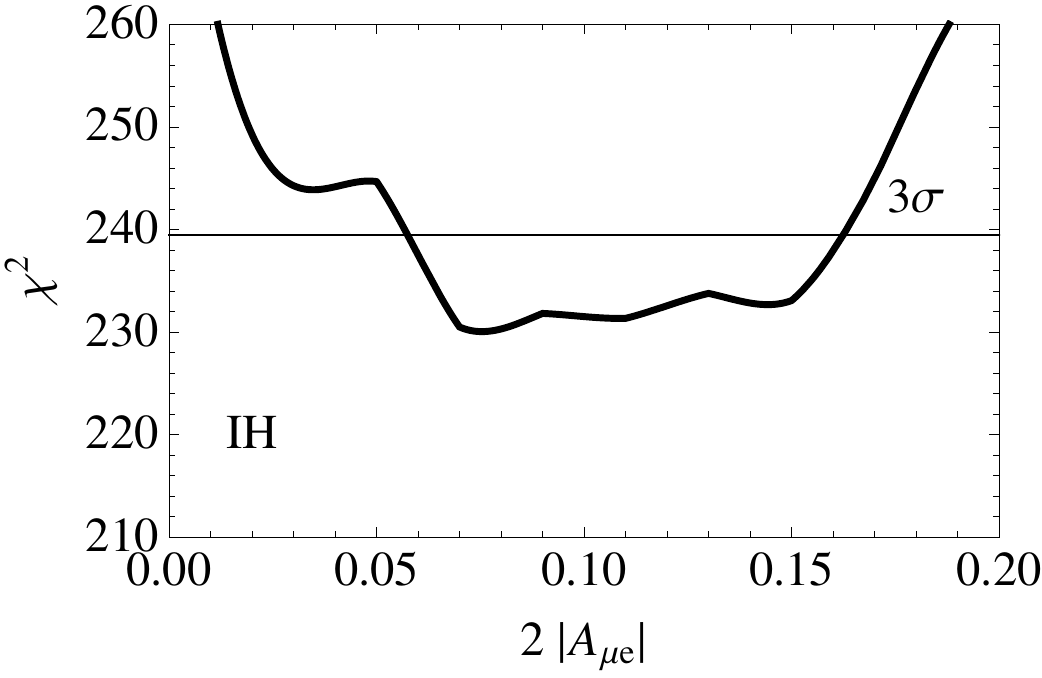}
\includegraphics[width=7cm]{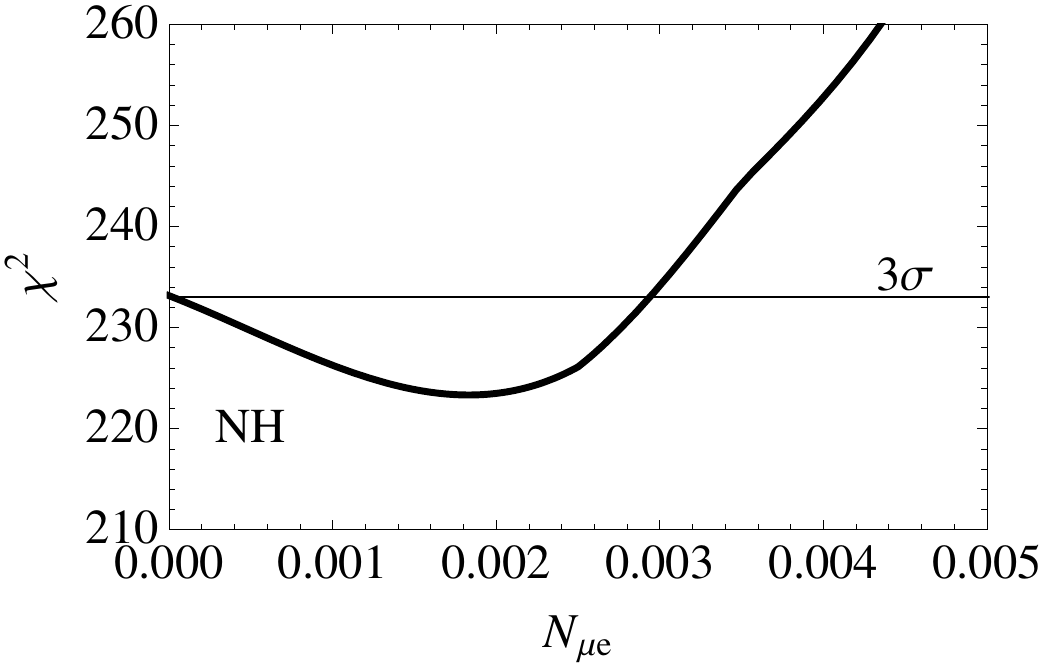}\includegraphics[width=7cm]{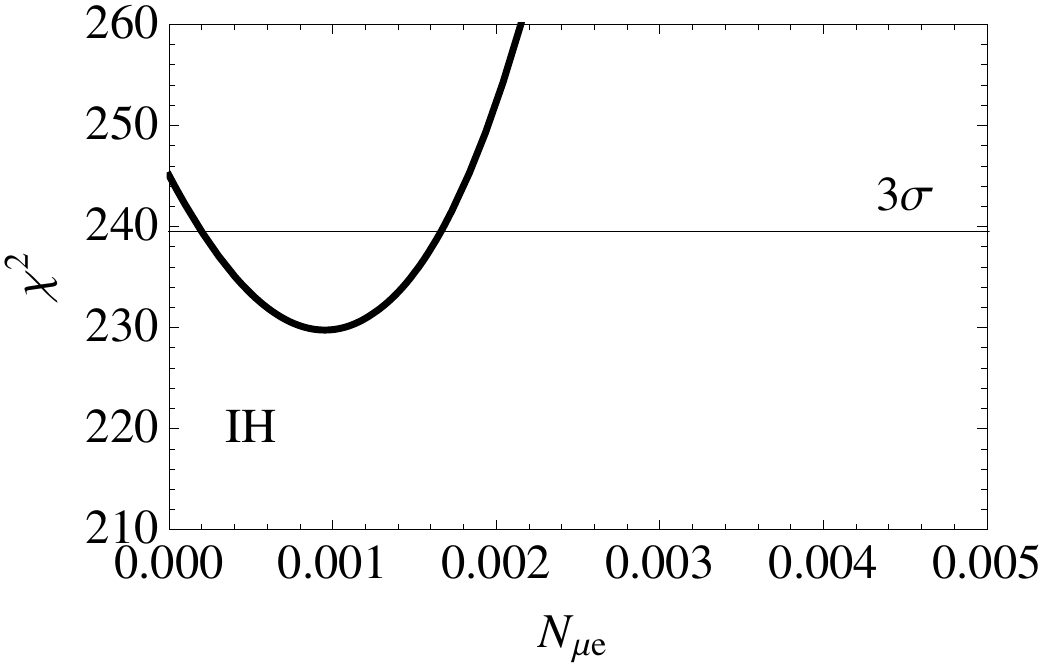}
\includegraphics[width=7cm]{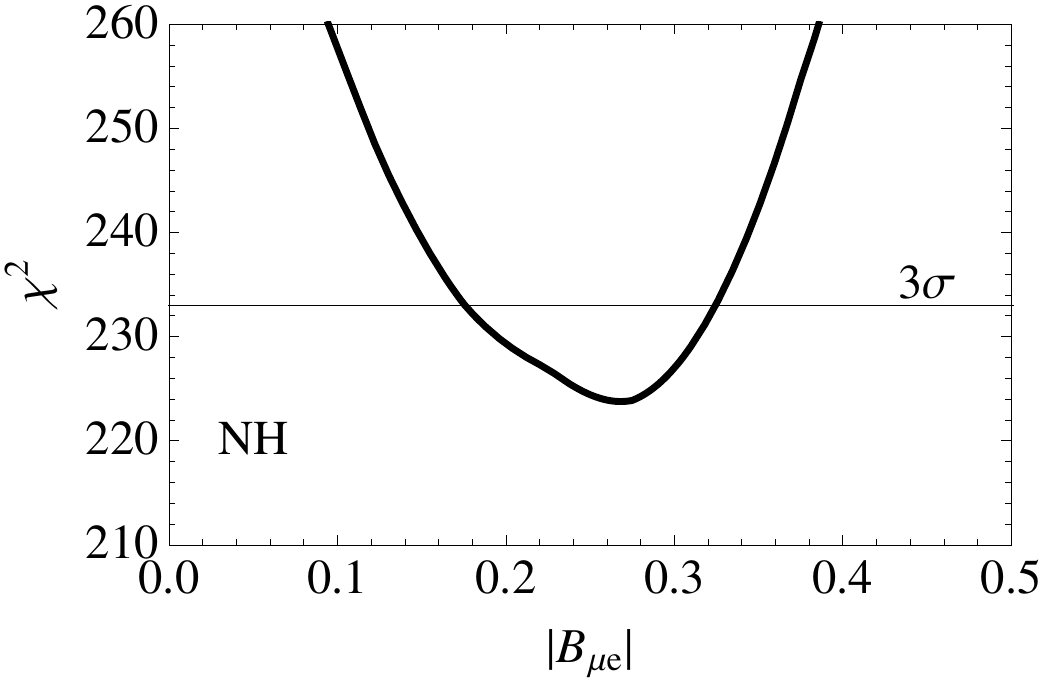}\includegraphics[width=7cm]{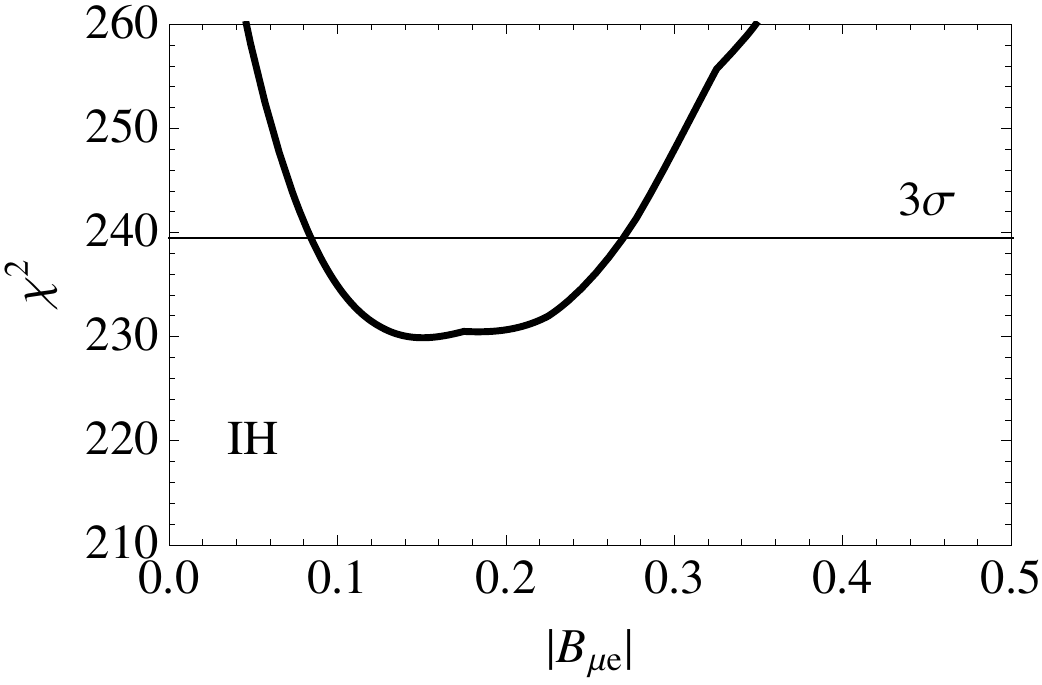}
\caption{ Minimum $\chi^2$ as a function of the physical combinations $A_{ee}$, $2 |A_{\mu e}|$, $N_{\mu e}$, $|B_{\mu e}|$ for the NH(left) and IH (right). The heavy masses are fixed to the KMS values of Table~2 for NH and to the GL values for IH. The band in the top plots corresponds to the recent Daya Bay 3$\sigma$ limit. }
\label{fig:phys}
\end{center}
\end{figure}

The projection of the $\chi^2$ as function of $A_{ee}$ is shown in the top plots of Fig.~\ref{fig:phys} for the NH and IH. It is interesting 
to see that the $3\sigma$ limits are more restrictive than the corresponding $3\sigma$ limits
of the recent Daya Bay  measurement, shown by the shaded region. 

The so-called golden-channel oscillation probability that will be measured in the T2K far detector (FD) and other future experiments is given, in the same approximation of one scale dominance (atmospheric), by
\begin{eqnarray}
\left. P_{\mu e}\right|_{\rm T2K-FD} \simeq N_{\mu e} - |A_{\mu e}|  \sin\left({\Delta_{atm} L \over 2}\right) \sin\left(\pm \phi + { \Delta_{atm} L \over 2}\right) +{\mathcal O}(\Delta_{solar}),\nonumber\\
\end{eqnarray}
where 
\begin{eqnarray}
N_{\mu e} &\equiv& 2 \left( |U_{e4}|^2 |U_{\mu 4}|^2+  |U_{e5}|^2 |U_{\mu 5}|^2 +\Re(U_{e4} U^*_{\mu 4} U^*_{e5} U_{\mu 5})\right),\nonumber\\
|A_{\mu e}| e^{\pm i \phi} &\equiv& \left\{\begin{array}{lrrr} 
4  ~U_{e3}^* U_{\mu 3} (U_{e1} U^*_{\mu 1}+U_{e2} U^*_{\mu 2}),  & & & {\rm NH}\\
4  ~U_{e1} U_{\mu 1}^* (U_{e3}^* U_{\mu 3}+U_{e2}^* U_{\mu 2}).  & & & {\rm IH}
\end{array}\right. \\
\end{eqnarray}
The $\pm$ refers to neutrinos and antineutrinos respectively. As explained in the appendix there is CP violation even 
  in the one scale dominance limit, e.g $\Delta_{21} \simeq 0$. The effect is however suppressed by the ammount of violation of unitarity of   the light-sector  (see appendix A), as expected on general grounds \cite{FernandezMartinez:2007ms}. 
  
\begin{figure}[htbp]
\begin{center}
\includegraphics[width=7cm]{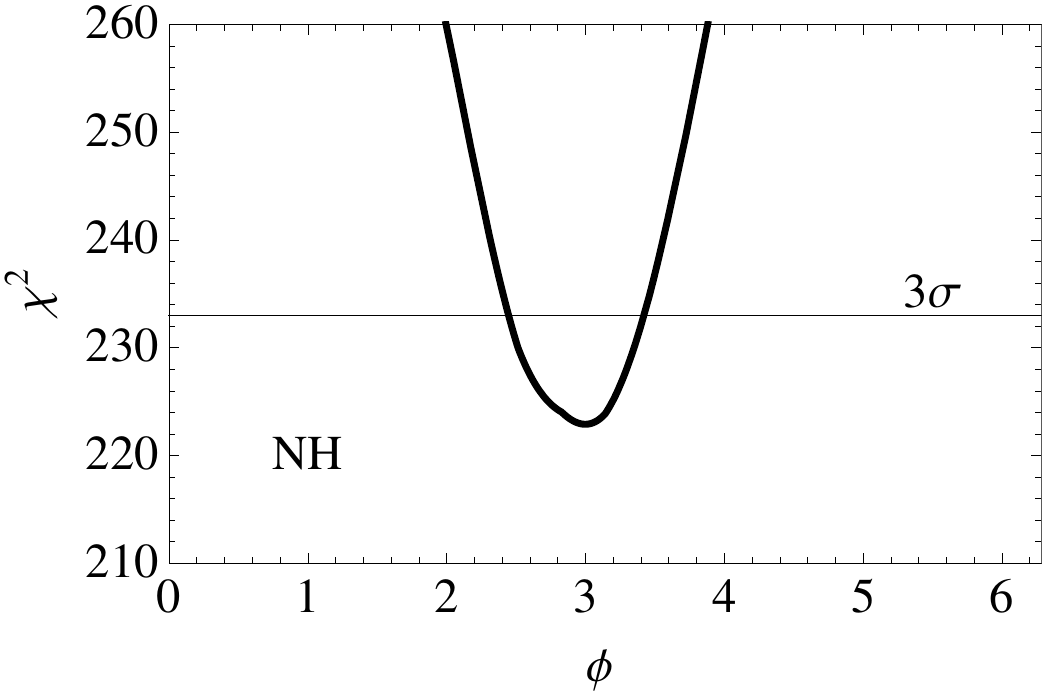}\includegraphics[width=7cm]{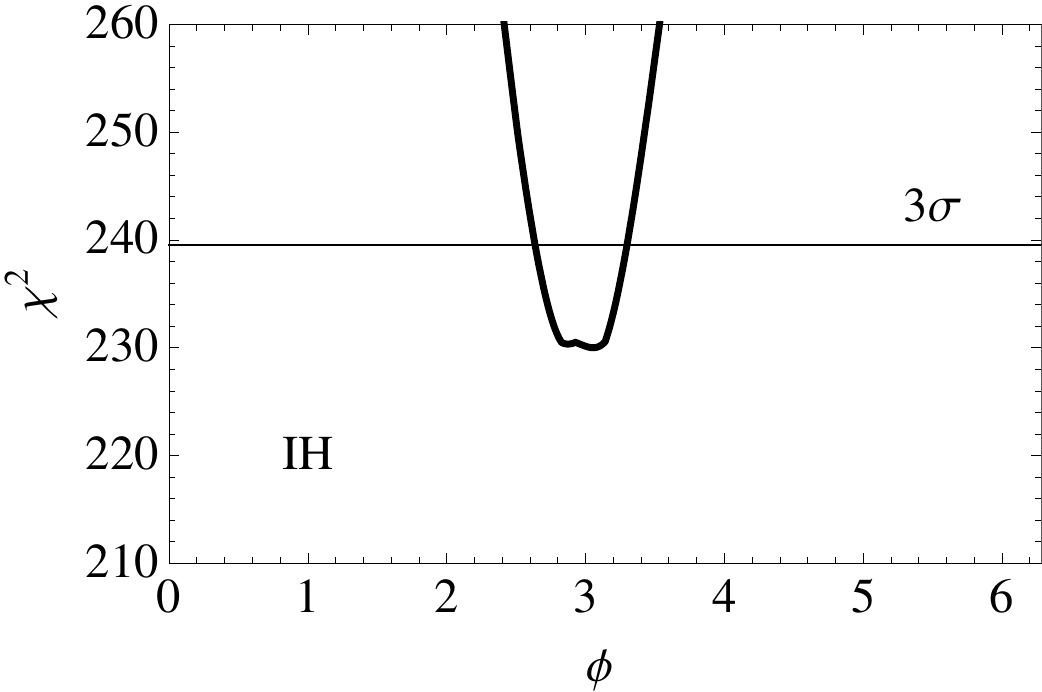}
\caption{$\chi^2$ as a function of the non-conventional phase $\phi$ for NH (left) and IH (right). The heavy masses are fixed to the KMS choice of  Table~2 for NH and to the GL choice for IH.}
\label{fig:chi2_phi}
\end{center}
\end{figure}

The minimum $\chi^2$ as a function $2 |A_{\mu e}|$  (which in the standard $3\nu$ 
scenario is roughly $A_{ee}$) and $N_{\mu e}$ (which would be zero in the $3\nu$ case) are shown in  the second and third row of Fig.~\ref{fig:phys}. The  constraint on the phase $\phi$ is shown in Fig.~\ref{fig:chi2_phi}. As expected, these non-conventional CP violating effects are  small (we can see that $\phi \sim \pi$ for NH and IH). 

The T2K experiment also has a near detector. However the baseline of 280~m is actually near the oscillation length corresponding to the  mass splittings of Table~\ref{tab:3+2}, so the approximation of averaged oscillations is not good in this case. A careful analysis would require to 
fit simultaneously the near and far detector oscillation probabilities. The appearance  probability in the near
detector  can be approximated by 
\begin{eqnarray}
P_{\mu e} & \simeq & 4 |U_{e 4}|^2 |U_{\mu 4}|^2 \sin^2 \left({M_1^2 L \over 4 E}\right)  +4 |U_{e 5}|^2 |U_{\mu 5}|^2 \sin^2 \left({M_2^2 L \over 4 E}\right)  \nonumber\\
&+& 8 |U_{e 4} U^*_{\mu 4} U^*_{e5} U_{\mu 5}| \sin \left({M_1^2 L \over 4 E}\right)     \sin \left({M_2^2 L \over 4 E}\right)   
\cos \left({ (M_2^2-M_1^2) L \over 4 E} \mp\phi_{45}\right),   \nonumber\\  
\end{eqnarray}
while the $\mu$ disappearance one, that can also be measured, is
\begin{eqnarray}
P_{\mu\mu} & \simeq &1-4 \left(1- |U_{\mu 4}|^2- |U_{\mu 5}|^2 \right) \left( |U_{\mu 4}|^2\sin^2\frac{M_1^2L}{4E}
+|U_{\mu 5}|^2 \sin^2\frac{M_2^2L}{4E} \right) \nonumber \\
&-&4 |U_{\mu 4}|^2|U_{\mu 5}|^2 \sin^2 \frac{(M_2^2-M_1^2)L}{4E}
\end{eqnarray}
This is also the good approximation for LSND and MiniBooNE. The corresponding mixing elements and the phase have been presented in the previous section.

 \begin{figure}[htbp]
\begin{center}
\includegraphics[width=7cm]{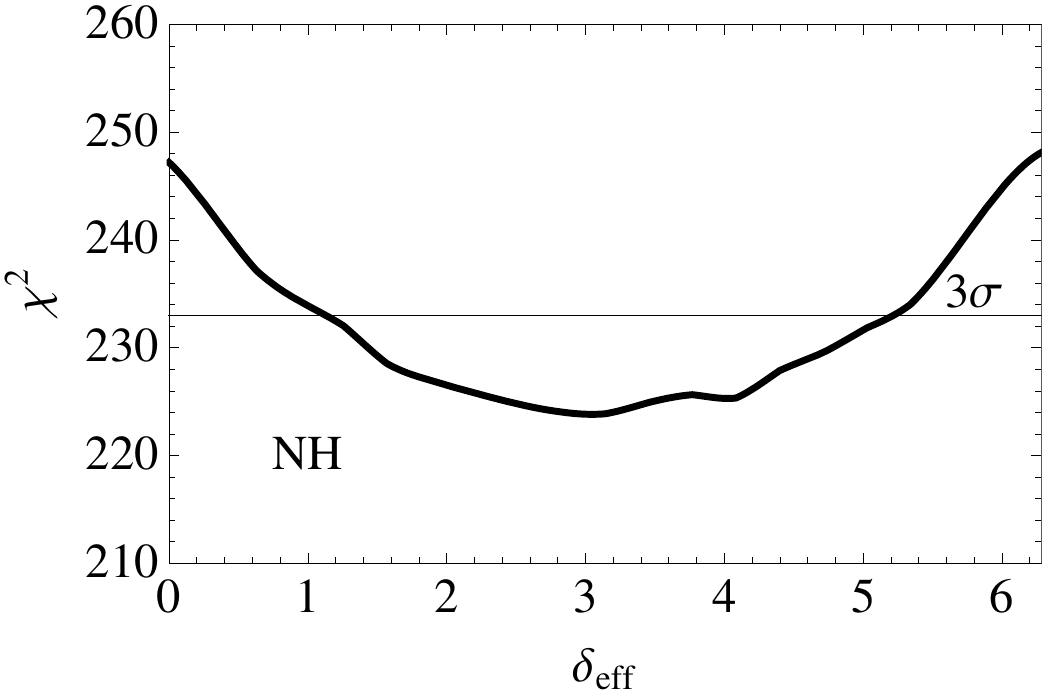}\includegraphics[width=7cm]{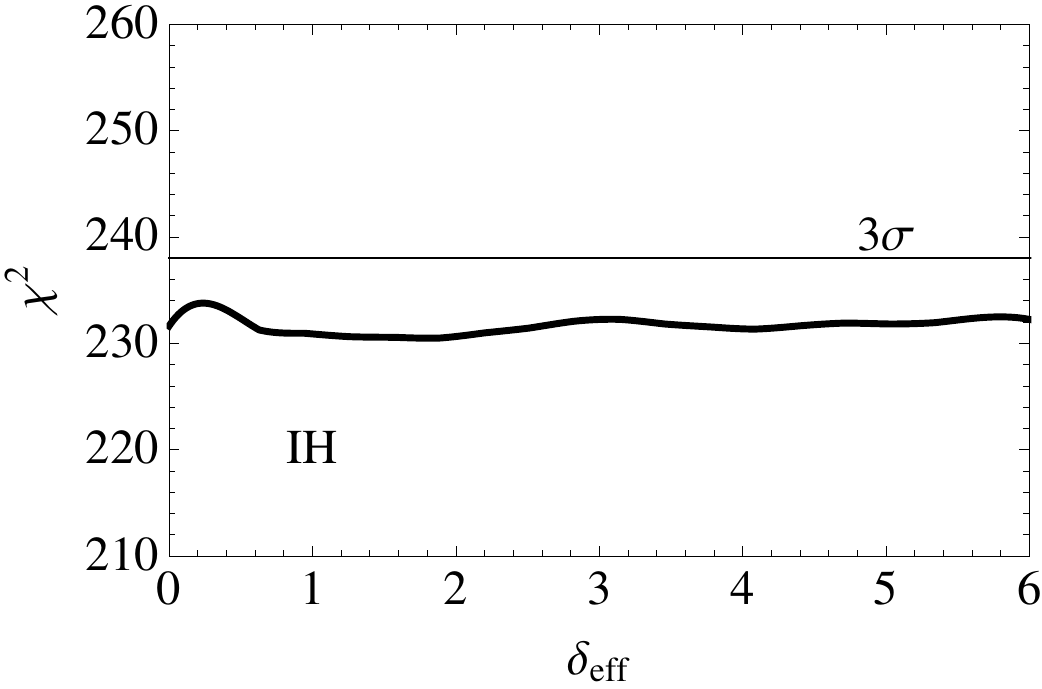}
\caption{ $\chi^2$ as a function of the effective $\delta_{\rm eff}$ for NH (left) and IH (right). The heavy masses are fixed to the KMS choice of  Table~2  for NH and to the GL choice for IH.}
\label{fig:chi2_cp3nu}
\end{center}
\end{figure}

\begin{figure}[htbp]
\begin{center}
\includegraphics[width=7cm]{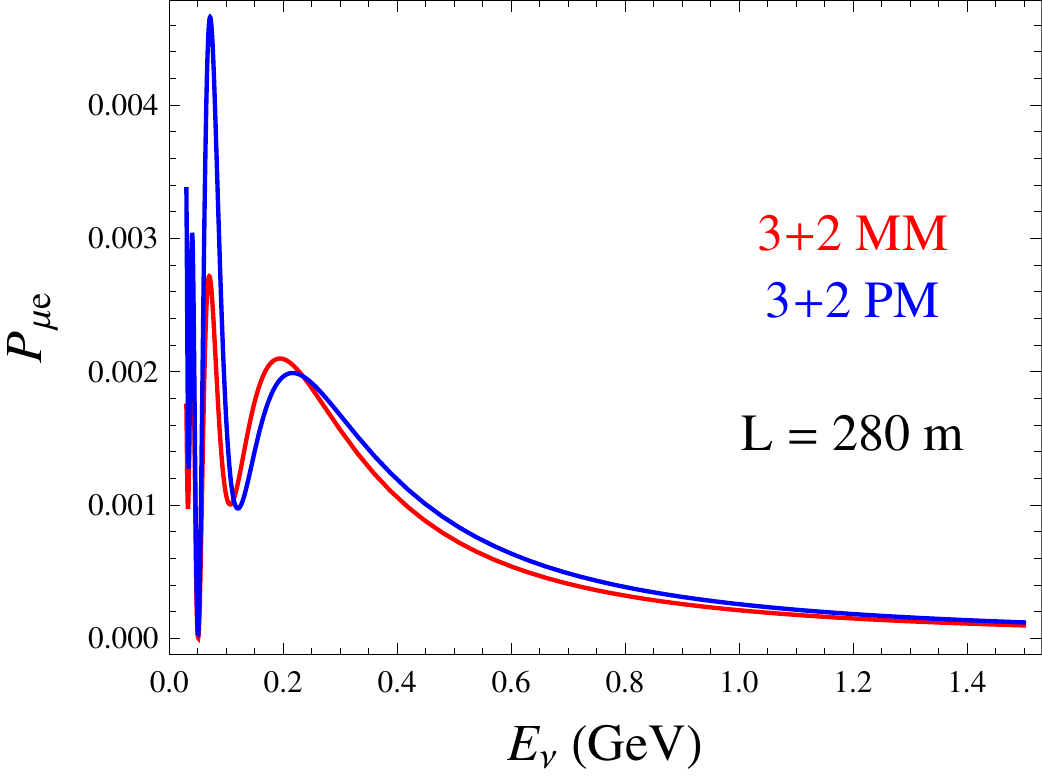}\includegraphics[width=7cm]{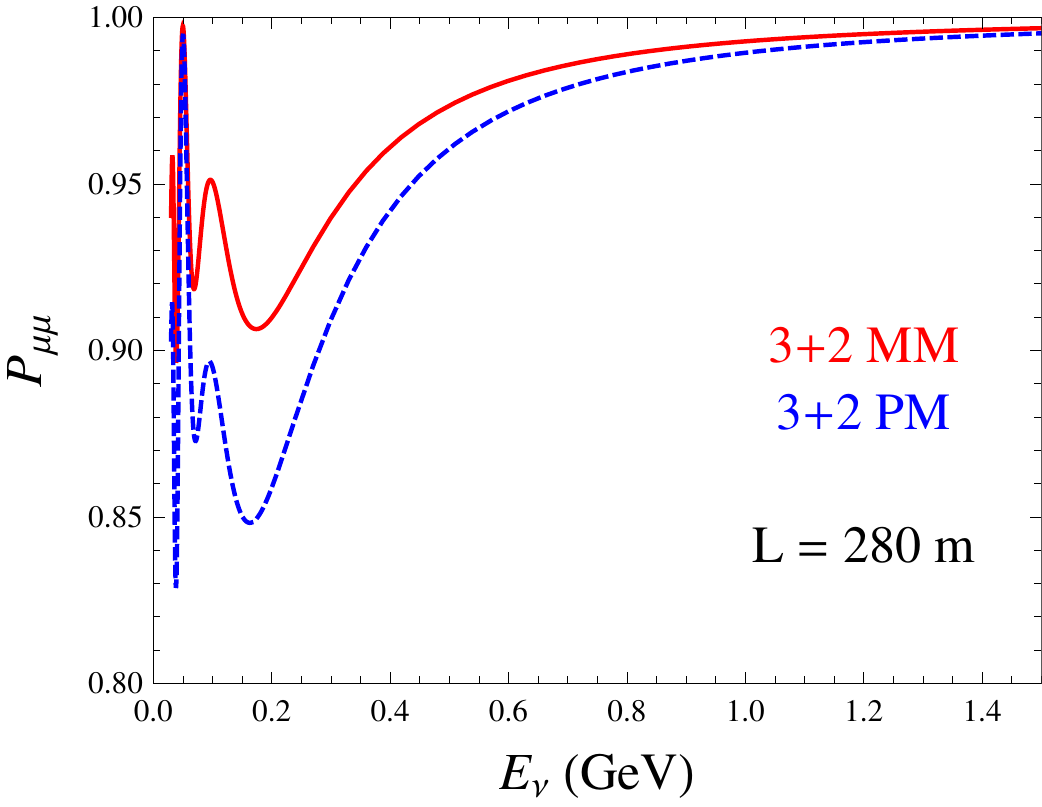}
\caption{Appearance probability $\mu e$ (left)  and $\mu$ disappearance (right) at a baseline $L=280$m (T2K near dectector) for the 3+2 MM best fit values for NH (solid line) and the 3+2 PM best fit values of \cite{Kopp:2011qd} (dashed line).  }
\label{fig:t2knd}
\end{center}
\end{figure}

In Fig.~\ref{fig:t2knd} we show the $P_{\mu e}$ and $P_{\mu\mu}$  probabilities for the  best fit of the 3+2 MM  (NH) at a baseline of $L=280$m and in the range of energies of the T2K beam. Clearly the T2K near detector could provide
very valuable information to further constrain the 3+2 MM.

Finally, in the longer term, the goal in future LBL experiments is to achieve sufficient precision to measure also the CP violation driven 
by the solar and atmospheric interference, ie. the only one present in the standard 3$\nu$ scenario. In this case, we cannot consider the limit of vanishing solar mass splitting.  A simultaneous expansion in $\Delta m^2 _{solar}$  and the small mixings  in eq.(\ref{fase}) gives the following approximate result
 \begin{eqnarray}
\left. P_{\mu e}\right|_{\rm T2K-FD} &\simeq& N_{\mu e} - |A_{\mu e}|  \sin\left({\Delta_{atm} L \over 2}\right) \sin\left(\pm \phi + { \Delta_{atm} L \over 2}\right) \nonumber\\
&+& |B_{\mu e}|  \cos \left(\pm\delta_{\rm eff} + {\Delta_{atm} L \over 2} \right) \sin {\Delta_{solar} L \over 2} \sin{\Delta_{atm} L \over 2}
\nonumber\\
&+&{\mathcal O}( \Delta^2_{solar},\epsilon \Delta_{solar}),
\end{eqnarray}
with 
\begin{eqnarray}
|B_{\mu e}| e^{\pm i \delta_{\rm eff}} &\equiv& \left\{\begin{array}{lrrr} 
8~U_{e2} U^*_{\mu 2} U^*_{e3} U_{\mu 3}  ,  & & & {\rm NH}\\
8~U_{e1} U^*_{\mu 1} U^*_{e3} U_{\mu 3}e^{i\pi}.  & & & {\rm IH}
\end{array}\right. \\
\end{eqnarray}

The minimum $\chi^2$ as a function of $|B_{e\mu}|$ is also shown in the last row of Fig.~\ref{fig:phys}. The constraint on the effective phase $\delta_{\rm eff}$ (that plays the same role as the standard CP phase in the $3\nu$ model) is shown in Figs.~\ref{fig:chi2_cp3nu}. We see that 
the range of values of $\delta_{\rm eff}$ in the 3+2 MM is poorly constrained: whilst a non-vanishing $\delta_{\rm eff}$ is expected for NH, all values are allowed  for IH.



\section{Conclusions and Outlook}
\label{sec:conclu}

We have studied a 3+2 minimal model with light sterile neutrinos, which corresponds to enlarging the Standard Model with two extra singlet Weyl fermions (often referred to as mini-seesaw models), and is arguably the simplest extension that can explain the observed neutrino masses. 
 The neutrino spectrum contains generically four massive  states and therefore can be represented as a 3+2 phenomenological model, but with constraints. Mixings and masses are correlated, as in seesaw models.
 
As it is well known, models with extra light sterile neutrinos in the $O$(eV) range do improve the global fits to neutrino oscillation data, because they can accommodate
the LSND signal (recent analyses can be found in \cite{Kopp:2011qd,Giunti:2011gz,Abazajian:2012ys}) and the new reactor anomaly, but 
there remains a significant tension in the $\mu$ sector between positive appearance signals and negative disappearance ones.  In the minimal model, it is possible to fix the same physical spectrum as found in the  3+2 phenomenological fits but, once the spectrum is fixed, the mixing parameters are severely constrained: only 4 angles and 3 phases can be  further tuned, while in the phenomenological 3+2 model, 9 angles and 5 phases enter in oscillation probabilities. The minimal model is therefore much more predictive.

We have evaluated to what extent the minimal model can fit global oscillation data, in spite of these constraints. We have presented a useful parametrization of seesaw models inspired by the Casas-Ibarra parametrization \cite{Casas:2001sr}, but  one that does not assume any expansion in the ratio of light and heavy masses. It reduces to Casas-Ibarra in the appropriate limit, but can be safely used even if corrections of higher order in the ratio of light to heavy masses are significant, as happens in our case. We believe this parametrization can be useful in other contexts. 

The results of the global fits show that the 3+2 minimal model outperforms the 3$\nu$ model in global fits by about a shift of 28 (20) units  of $\chi^2$ for NH (IH). Previous estimates based in Casas-Ibarra parametrization \cite{Donini:2011jh,deGouvea:2011zz,Fan:2012ca} indicated that the electron mixings to the heavy states would be too small for the NH. We arrive to a different conclusion, due in part to the large value of the recently measured reactor angle and to the fact that higher order corrections in the heavy-light mixings are significant. The minimal model for the NH predicts however significantly larger tau mixings to the heavy states. 

Compared to the 3+2 phenomenological fits, we find that the minimal models, in spite of the constraints, give similarly good fits, especially in the case of the NH.

Experimental evidence of extra light sterile neutrinos is slowing accumulating from oscillations experiments and cosmology. Unfortunately, 
the significant signal from LSND has not yet been confirmed at a similar level of confidence by any other experiment. If such a signal would be confirmed in the future, it might be explained by just adding two Weyl singlet fermions to the Standard Model.

\acknowledgments

We wish to thank Joaquim Kopp for useful discussions. 
This work was partially supported by the Spanish Ministry for Education and Science projects  FPA2007-60323, FPA2009-09017 and FPA2011-29678; the Consolider-Ingenio CUP (CSD2008-00037) and CPAN (CSC2007-00042); the Generalitat Valenciana (PROMETEO/2009/116);   the European projects EURONU (CE212372), LAGUNA (Project Number 212343) and the ITN INVISIBLES (Marie Curie Actions, PITN- GA-2011-289442).

\appendix
\section{Appendix}
\label{sec:appendix}
We gather here some useful expressions for the oscillation probabilities in 3+2 models. 

For values of $E/L$ such that the heavier splittings are in the averaged-oscillation regime, the oscillation probability  is 
well approximated by the following expression
\begin{eqnarray}
\label{fase}
P_{\alpha\beta}&=& \delta_{\alpha\beta}( 1-2  \epsilon_{\alpha\alpha})  
+2\left(|\epsilon_{\alpha\beta}|^2-|U_{\alpha4}U_{\beta 4}^*U_{\alpha5}^*U_{\beta 5}|\cos \phi_{45}\right)
\nonumber\\
&-&4 |U_{\alpha2}U_{\beta 2}^*|\left(\delta_{\alpha\beta}-|U_{\alpha2}U_{\beta 2}^*|\right) \sin^2\frac{\Delta_{21} L}{2}-
4|U_{\alpha3}U_{\beta 3}^*|\left(\delta_{\alpha\beta}-|U_{\alpha3}U_{\beta 3}^*|\right)\sin^2\frac{\Delta_{31} L}{2}
\nonumber\\
\nonumber\\
&+&8|U_{\alpha 2}U_{\beta 2}^*U_{\alpha 3}^*U_{\beta 3}|\cos\left( \pm\phi_{23}-\frac{\Delta_{32}L}{2}\right)\sin\frac{\Delta_{21} L}{2}\sin\frac{\Delta_{31} L}{2}\nonumber\\
&+&4|U_{\alpha3}U_{\beta 3}^*\epsilon_{\alpha\beta} |\sin\left(\pm  \phi_{3}+\frac{\Delta_{31} L}{2}\right)\sin\frac{\Delta_{31}L}{2}
\nonumber\\
&+&4|U_{\alpha 2}U_{\beta 2}^*\epsilon_{\alpha\beta}|\sin\left(\pm \phi_{2}+\frac{\Delta_{21} L}{2}\right)\sin\frac{\Delta_{21}L}{2}
\nonumber\\
\end{eqnarray}
with
\begin{eqnarray}
\epsilon_{\alpha\beta}\equiv \delta_{\alpha\beta}-\sum_{i=1}^3 U_{\alpha i}^*U_{\beta i}\;\;\;
 \phi_{ij}\equiv \arg\{U_{\alpha i}U_{\beta i}^*U_{\alpha j}^*U_{\beta j}\},\;\;\;  \phi_{i}\equiv \arg\{U_{\alpha i}U_{\beta i}^*\epsilon_{\alpha\beta}\} .
\end{eqnarray}
The $\epsilon$ terms measure the violation of unitarity in the light 123 sector and are therefore the terms that are non-standard. The $3\nu$ probabilities are recovered in the limit $\epsilon, U_{\alpha 4}, U_{\alpha 5}  \rightarrow 0$. 

As expected  there is CP violation even in the one scale dominance limit, e.g $\Delta_{21} \simeq 0$. The effect is however suppressed in light-sector non-unitarity, as expected on general grounds \cite{FernandezMartinez:2007ms}. The corresponding CP violating terms (last two terms in eq.~(\ref{fase})) are indeed suppressed in $O(\epsilon)$. 

Let us rewrite Eq.~(\ref{fase}) for $\alpha\neq\beta$ expanding over the solar oscillation frequency. For the NH case
we obtain:

\bea
\label{faseNH}
P_{\alpha\beta}^{NH}&=&2\left(|U_{\alpha4}|^2|U_{\beta 4}|^2+|U_{\alpha5}|^2|U_{\beta 5}|^2+|U_{\alpha4}U_{\beta 4}^*U_{\alpha5}^*U_{\beta 5}|\cos \phi_{45}\right)
\nonumber\\
&-&4|U_{\alpha3}U_{\beta 3}^*\left(U_{\alpha1}^*U_{\beta 1}+U_{\alpha2}^*U_{\beta 2}\right)|\sin\left(\pm \phi+\frac{\Delta_{31}L}{2}\right)\sin\frac{\Delta_{31} L}{2}
\nonumber\\
&+&4|U_{\alpha2}U_{\beta 2}^*U_{\alpha3}^*U_{\beta 3}|\cos\left(\pm \phi_{23}-\frac{\Delta_{31}L}{2}\right)\left(\Delta_{21}L\right)\sin\frac{\Delta_{31} L}{2}
\nonumber\\
&\pm& 2|U_{\alpha2}U_{\beta 2}^*\epsilon_{\alpha\beta}|\sin\phi_{2}(\Delta_{21}L)+\mathcal{O}\left(\left(\Delta_{21}L\right)^2\right)
\eea

where 
\be
\phi=\text{arg}\{U_{\alpha3}U_{\beta 3}^*\left(U_{\alpha1}^*U_{\beta 1}+U_{\alpha2}^*U_{\beta 2}\right)\}\nonumber
\ee
and the phases $\phi_{i}$ and $\phi_{ij}$ have been defined above. As expected, the appearance probability 
in the IH case can be obtained just doing $123\rightarrow231$ in the above equations:
\be
\label{faseIH}
P_{\alpha\beta}^{IH}=P_{\alpha\beta}^{NH}\left(123\rightarrow231\right)
\ee
Of course, the same symmetry between NH and IH is shown by the disappearance (and appearance) probabilities
presented in Sec.~\ref{sec:pheno}. In this way we are expressing the probabilities 
as a function of the solar and atmospheric mass differences since the transformation $123\rightarrow231$ 
corresponds to the following mapping:  
\bea
\Delta_{31}&=&\Delta_{atm} \rightarrow -\Delta_{21}=-\Delta_{atm}\nonumber\\
\Delta_{21}&=&\Delta_{sol} \rightarrow \Delta_{32}=\Delta_{sol} 
\eea
Finally, note that no expansion over $\epsilon_{\alpha\beta}$ has been considered here.

\bibliographystyle{h-elsevier}
\bibliography{biblio}

\end{document}